\documentclass[10pt,twocolumn,letterpaper]{article}
\usepackage{iccv}              
\usepackage{placeins} 
\usepackage{float}
\usepackage{algorithm}
\usepackage{algorithmic}
\usepackage{tabularx}
\usepackage{graphicx}
\usepackage{multirow}

\newcommand{\todo}[1]{{\color{red}#1}}

\newcommand\tofix[1]{\noindent{\textcolor{magenta}{#1}}}

\newcommand\deletion[1]{}

\definecolor{iccvblue}{rgb}{0.21,0.49,0.74}
\usepackage[pagebackref,breaklinks,colorlinks,allcolors=iccvblue]{hyperref}
\usepackage{colortbl}

\def\paperID{9120} 
\def\confName{ICCV}
\def\confYear{2025}

\newcommand{\X}{ContraGS\xspace}
\title{\X: \tofix{Co}debook-\tofix{Con}densed and \tofix{Tra}inable \tofix{G}aussian \tofix{S}platting for Fast, Memory-Efficient Reconstruction}

\author{Sankeerth Durvasula$^1\textsuperscript{\textdagger}$, \hspace{0.1em}
Sharanshangar Muhunthan$^1$, \hspace{0.1em}
Zain Moustafa$^1$, \hspace{0.1em}
Richard Chen$^1$, \hspace{0.1em}
Ruofan Liang$^1$ \\ 
Yushi Guan$^1$, \hspace{0.3em}
Nilesh Ahuja$^2$, \hspace{0.3em}
Nilesh Jain$^2$, \hspace{0.3em}
Selvakumar Panneer$^2$, \hspace{0.3em}
Nandita Vijaykumar$^1$\\
$^1$University of Toronto \hspace{1.6em} $^2$Intel \\
\texttt{\{sankeerth,zain,ruofan,guanyushi,nandita\}@cs.toronto.edu}\\
\texttt{\{shangar.muhunthan,riixardo.chen\}@mail.utoronto.ca} \\
\texttt{\{nilesh.ahuja,nilesh.jain,selvakumar.panneer\}@intel.com} \\
}

\begin{document}

\setlength{\tabcolsep}{3pt}
\setlength{\fboxrule}{.1pt}
\renewcommand{\arraystretch}{1}
\twocolumn[{%
\renewcommand\twocolumn[1][]{#1}%
\maketitle
\begin{center}
    \centering
    \captionsetup{type=figure}
    \includegraphics[width=0.95\textwidth]{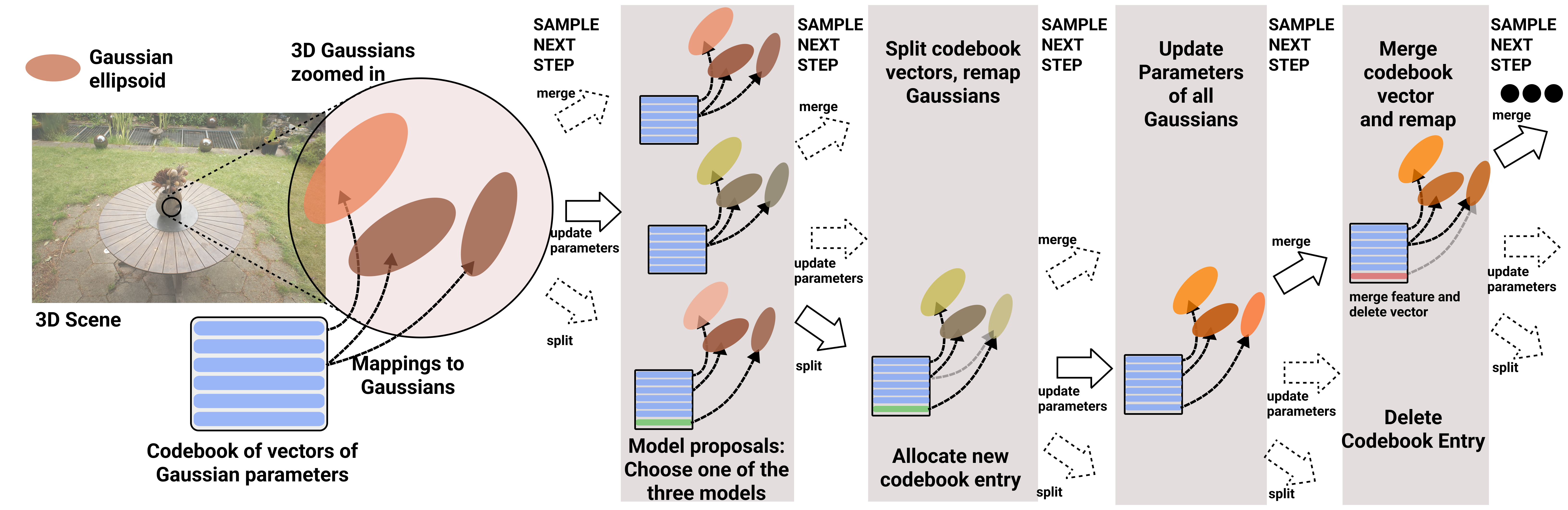}
    \captionof{figure}{\X: A 3D scene is modeled using 3D Gaussian Splatting~\cite{kerbl20233d}. A codebook stores a compressed representation of the 3DGS scene as common vectors of 3DG parameters. \X enables learning parameters of a codebook-compressed 3DG scene by posing the parameter estimation as a Bayesian inference problem. \X{} proposes split/merge/update on each codebook vector to explore the state space of codebook-compressed models.}
    \label{fig:mainfig}    
\end{center}
}]


\begin{abstract}

3D Gaussian Splatting (3DGS) is a state-of-art technique to model real-world scenes with high quality and real-time rendering.
Typically, a higher quality representation can be achieved by using a large number of 3D Gaussians.
However, using large 3D Gaussian counts significantly increases the GPU device memory for storing model parameters. A large model thus requires powerful GPUs with high memory capacities for training and has slower training/rendering latencies due to the inefficiencies of memory access and data movement. 
In this work, we introduce \X, a method to enable training directly on compressed 3DGS representations without reducing the Gaussian Counts, and thus with a little loss in model quality. \X leverages codebooks to compactly store a set of Gaussian parameter vectors throughout the training process, thereby significantly reducing memory consumption. 
While codebooks have been demonstrated to be highly effective at compressing fully trained 3DGS models, directly \emph{training} using codebook representations is an unsolved challenge. \X solves the problem of learning non-differentiable parameters in codebook-compressed representations by posing parameter estimation as a Bayesian inference problem. To this end, \X provides a framework that effectively uses MCMC sampling to sample over a posterior distribution of these compressed representations. With \X, we demonstrate that \X significantly reduces the peak memory during training (on average $3.49\times$) and accelerated training and rendering ($1.36\times$ and $1.88\times$ on average, respectively), while retraining close to state-of-art quality.


\end{abstract}


\vspace{-0.4cm}
\section{Introduction}
\label{sec:introduction}
3D Gaussian Splatting~\cite{kerbl20233d} (3DGS) is a state-of-art technique for modeling real-world scenes with high quality and real-time rendering. It is a powerful scene representation method applicable to various tasks, including 3D/4D reconstruction~\cite{4dsplat,deformable3d,gaussiansinthewild,gaussianflow,atomgs,splatfield,gs-lrm}, simultaneous localization and mapping (SLAM)~\cite{gs-slam,cg-slam,dg-slam,fgs-slam,splatmap}, 3D object generation~\cite{dreamgaussian,cat3d,brightdreamer,generativeobjectinsertion} 3D editing~\cite{gsedit, gaussianeditor,gaussctrl,direct-dge,dragyourgaussian,gaussiangrouping,3dsceneeditor} 
simulation~\cite{physgaussian,physicallyembodied-gs,robogsim,gaussianproperty}, representing detailed human avatars~\cite{humansplat-kocabas,humangaussiansplatting-moreau,3dgs-avatar,splattingavatar,pardy-human,haha,exavatar,gavatars}
, etc. 
3DGS represents scenes with a collection of 3D Gaussians associated with a view-dependent emitted radiance (color). 
The parameters defining the position, shape and color of each Gaussian are obtained by minimizing a photometric loss using gradient descent. 

\begin{figure}[!t]
    \centering
    \includegraphics[width=0.8\linewidth]{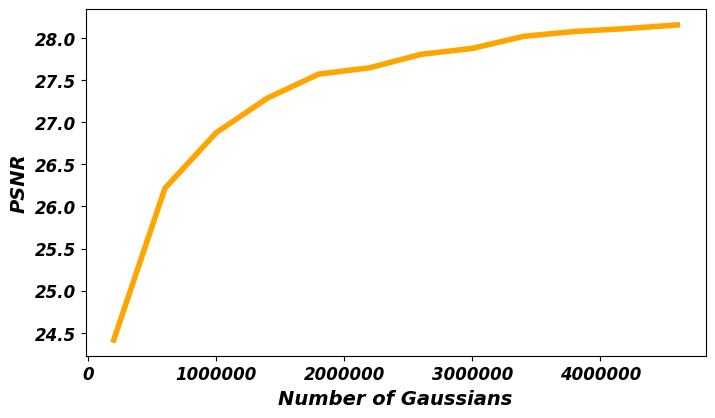}
    \vspace{-0.2cm}
    \caption{Impact of number of Gaussians on representation quality for the ``Truck'' scene of T\&T~\cite{tandt}, using 3DGS-MCMC~\cite{3dgsmcmc}}
    \label{fig:psnr_num3dg_intro}
    \vspace{-0.5cm}
\end{figure}

\noindent Learning very high quality scene representations with 3DGS, especially for complex scenes, typically requires a large number of 3D Gaussians.
Fig.~\ref{fig:psnr_num3dg_intro} shows the reconstruction quality (PSNR) vs. the number of Gaussians used to represent the ``Truck'' scene~\cite{tandt}, using the state-of-art approach 3DGS-MCMC~\cite{3dgsmcmc}. $2$ million Gaussians achieves a PSNR of $27.64$, whereas $4$ million Gaussians achieves a PSNR of $28.15$.
Using more Gaussians, however, incurs much larger memory overheads during both training/inference and for storage. 
For example, representing the Truck scene~\cite{tandt} with Gaussians requires over $2$ GB of memory for the model alone. Efficiently learning scenes with high Gaussian counts thus necessitates a powerful GPU with large memory capacity. The peak memory needed for training can be as much as $4\times$ the model size in order to save intermediate values during rendering and gradient computation.
In addition to requiring GPUs with large memory capacities, this also significantly slows down the training process due to the large data movement overheads. Thus, training complex scenes with 3DGS is challenging on environments such as mobile devices, web browsers, and other resource-constrained platforms.

\noindent A range of recent works~\cite{c3dgs, lightgs, rdgaussian, compact3d, pup3dgs} propose post-training compression methods to reduce the amount of memory required to store a 3DGS model. However, these works still incur the full memory cost \emph{during} training. Other prior works~\cite{turbogs, scaffolds, eagles, speedysplat} reduce model size during training by pruning redundant Gaussians. However, this results in lower quality models than those that can be achieved with a larger number of Gaussians (Section~\ref{sec:quality_metric_results}).

\noindent In this work, we aim to enable high-speed memory-efficient training and rendering without sacrificing the quality of a 3DGS scene with high Gaussian counts. 
To this end, we leverage codebook representations~\cite{c3dgs, compactgs, lightgs} that can significantly reduce the memory required to store 3DGS parameters. Codebooks reduce the model's memory footprint by storing a small set of vectors containing parameters shared among Gaussians, instead of storing an independent set of parameters for each Gaussian. Each Gaussian maps to one vector in the codebook.
Codebooks have been demonstrated to be highly effective at compressing fully trained 3DGS models~\cite{lightgs, c3dgs, compactgs} as a \emph{post-training} optimization. However, leveraging codebook compression \emph{during training} is an unsolved challenge. The parameters in the codebook compressed representation are (a) the codebook vectors and (b) the mapping between Gaussians and their codebook vectors. 
These non-differentiable indexes that map Gaussians to codebook vectors cannot be learnt by SGD.




\noindent In this work, we solve the problem of directly learning on codebook-compressed representations by using MCMC sampling to sample over a posterior distribution of these compressed representations. 
Posing the parameter estimation as Bayesian inference over a posterior distribution~\cite{3dgsmcmc}, instead of an optimization problem, offers the opportunity to estimate the non-differentiable compressed-3DGS parameters.
However, using Stochastic Gradient Langevin Dynamics (SGLD), a method for Bayesian inference as proposed by prior work, 3DGS-MCMC~\cite{3dgsmcmc} will not work for codebook-compressed representations. This is because SGLD also requires the differentiability of the log-posterior distribution with respect to estimated parameters. 

\noindent We propose \X, a method for learning codebook-compressed representations via Bayesian inference. 
We use the Metropolis-Hastings algorithm, a Markov Chain Monte Carlo sampling method, to sample from the posterior distribution.  
An overview of \X is shown in Fig.~\ref{fig:mainfig}. Starting from an initialized set of mappings between the Gaussians and the codebook, \X defines a proposal distribution 
that generates new candidate compressed 3DGS-models with a different set of Gaussian-to-Codebook mappings.
This new set of models is obtained by either splitting/cloning a set of codebook vectors into two vectors or merging pairs of codebook vectors into one vector.
Gaussians are remapped to the new codebook vectors generated by splitting/merging. 
Different sequences of splitting and merging operations will generate all possible codebook-compressed model representations. 
Some steps may simply update the values of the codebook vectors without affecting the mapping.
Thus, proposals drawn from this distribution enable exploration of the complete state space of codebook-compressed models.

\noindent We demonstrate that \X enables high-speed memory-efficient training without sacrificing the model quality significantly.
With 2 million Gaussians used to represent the scene, \X reduces the peak model memory by $3.49\times$ on average during training. This accelerates training by $1.36\times$ on average and increases the final model's rendering FPS by $1.88\times$ on average. Our contributions can be listed as follows:
\begin{itemize}
    \item This is the first proposed method that enables 3DGS training directly on compressed representations.
    \item We propose a novel mathematical framework to jointly learn codebook vectors and their mappings to model parameters, achieving efficient compression \emph{during} training. We formulate codebook compression as an MCMC sampling problem over a posterior distribution. We use this approach to learn non-differentiable codebook indices by exploring the state space of possible codebook mappings during MCMC sampling. 
    This framework can be applied to learn other codebook-compressed point-based 3D reconstruction methods to enable fast, memory efficient training such as RadiantFoam~\cite{radiantfoam}, ADOP~\cite{adop}, Deformable Beta Splatting~\cite{dbsplatting} and LinPrim~\cite{linprim}.
    \item We show how adopting a Bayesian inference formulation over a posterior allows incentivising smaller codebooks sizes for higher compression. 
    \item We demonstrate that \X significantly reduces peak memory during training while retaining close to state-of-art quality. For any memory capacity constraint (i.e, peak memory utilization), \X achieves the highest representation quality compared to prior works.





\end{itemize}

\section{Background}
\label{sec:background}

\subsection{3DGS Scene Representation}
\label{sec:3dgs_background}

Point-based scene representation techniques, such as 3DGS~\cite{kerbl20233d}, represent 3D scenes with a collection of anisotropic 3D Gaussian densities in space. The $i^{th}$ Gaussian distribution $G_i$ is given by:
\begin{equation}
\label{eq:gaussiandensity}
G_i(\boldsymbol{r}) = o_i  \mathrm{exp}\left\{ -\frac{1}{2} (\boldsymbol{r}-\boldsymbol{\mu}_i)^T\Sigma_i^{-1}(\boldsymbol{r}-\boldsymbol{\mu}_i)\right\}    
\end{equation}
Where $\boldsymbol{r}$ is a 3D position, $o_i$ is the opacity, $\mu_i$ is the location of its center, and $\Sigma_i$ is the covariance. $\Sigma_i$ is expressed as:
\begin{equation}
\label{eq:sigma}    
\Sigma = R(\boldsymbol{q})SS^TR(\boldsymbol{q})^T
\end{equation}
Where $R(\boldsymbol{q})$ represents the rotation matrix described by the quaternion $\boldsymbol{q}$, and $S$ is a diagonal matrix corresponding to principal axis scales. Rendering the  scene for a camera pose $\mathcal{P}$ involves determining the color $\boldsymbol{C}$ of each pixel $x$, the scene comprising Gaussians can be rendered using the following equations:
\begin{equation}
\label{eq:3dgcolor}
\boldsymbol{C}(\boldsymbol{x}) = \sum_{k=1}^N\alpha_k(\boldsymbol{x})\boldsymbol{c}_k \prod_{i=1}^{k-1}(1-\alpha_i(\boldsymbol{x}))
\end{equation}
Here, $\alpha_i$ is the 2D Gaussian projected from the 3D Gaussian density (Eq.~\ref{eq:gaussiandensity}) onto the camera plane. $\boldsymbol{c}_i$ is the color of the 3D Gaussian as seen in the viewing direction.

\noindent From a set of $M$ images $I_{\mathcal{P}_1}, I_{\mathcal{P}_2}, I_{\mathcal{P}_3}, ..., I_{\mathcal{P}_M}$ of a 3D scene taken from viewing directions $\mathcal{P}_1, \mathcal{P}_2, \mathcal{P}_3, ..., \mathcal{P}_M$, 3D reconstruction aims to determine the 3D Gaussians parameters that capture the 3D scene. Parameters are determined by minimizing a loss function $\mathcal{L}_{\text{recon}}$, which measures the accuracy of reconstruction. $\mathcal{L}_{\text{recon}}$ is expressed as:
\begin{equation}
\label{eq:recon_loss}
\mathcal{L}_{\text{recon}} = \left(1-\lambda_{\text{ssim}}\right) \mathcal{L}_1 + \lambda_{\text{ssim}}\mathcal{L}_{\text{SSIM}}
\end{equation}
Here, $\mathcal{L}_1$ and $\mathcal{L}_{\text{SSIM}}$ are the averaged L1 and Structural Similarity (SSIM) losses between the rendered and ground-truth images. 
The parameters that minimize the loss function are determined using gradient descent.


\subsection{Codebooks to Store 3DG Parameters}
\label{sec:background_codebooks}

3D scenes typically consist of many Gaussians with similar sets of parameters. \deletion{For example, in Fig.~\ref{fig:common_sceneparts}, (image, memory associations diagram), sections of the scene consist of similar characteristics such as color, principle axis scales, etc.} The memory needed for storing these redundant Gaussian parameters can be reduced by using codebooks. A codebook is a small set of vectors storing 3D Gaussian parameters. Each Gaussian is mapped to a vector in the codebook. A Gaussian's parameters can be obtained by looking up the mapped vector in the codebook. 
\begin{figure}[!htb]
    \vspace{-0.4cm}
    \centering
    \includegraphics[width=0.9\linewidth]{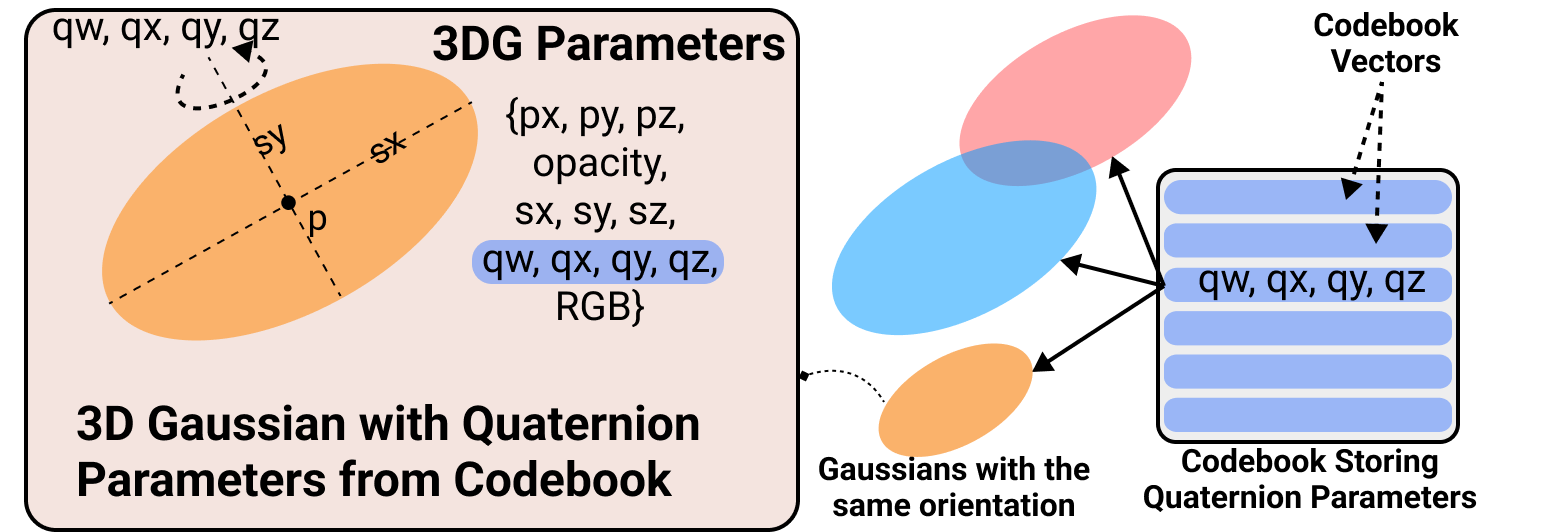}
    \caption{Gaussians mapped to one vector in a codebook of quaternions (right hand side). One Gaussian among them, whose quaternions are derived from the codebook vector is shown on the left.}
    \label{fig:codebook}
    \vspace{-0.4cm}
\end{figure}

C3DGS~\cite{c3dgs} uses codebooks to compress 3DGS scene. It uses one codebook to store spherical harmoics coefficients, and another codebook to store covariance parameters (rotation quaternion and principal axis scales) as shown in Fig.~\ref{fig:codebook}. Each Gaussian is mapped to one vector in the SH feature codebook, and one vector in the covariance codebook. 
Note that although we adopt a similar codebook indexing scheme as C3DGS~\cite{c3dgs}, our method does not restrict us from choosing this specific way of choosing codebooks.

\subsection{3DGS Training as MCMC Sampling}
\label{sec:3dgs_mcmc}

3DGS-MCMC~\cite{3dgsmcmc} demonstrated that the  training process of 3DGS can be interpreted as performing bayesian inference over the probability distribution function given by:
\begin{equation}
\label{eq:mcmcobjective}
p\left(G\right) = \frac{1}{Z} \mathrm{exp}\left(  -\mathcal{L}_{\text{recon}}(G) \right)
\end{equation}
Here, $G$ is the set of parameters of all Gaussians in the scene, $Z$ is\deletion{the partition function} a normalizing constant to the probability distribution.
$\mathcal{L}_{\text{recon}}$ is the reconstruction loss given by Eq.~\ref{eq:recon_loss}, written as a function over Gaussian parameter $G$. Bayesian inference is performed by sampling a set of Gaussian parameter values $G_{\text{recon}}$ from the distribution $p$ (denoted as $G_{\text{recon}}  \sim p$). Sampling from this high dimensional distribution function can be performed using Markov Chain Monte Carlo (MCMC) sampling.

\subsubsection{MCMC Sampling via Metropolis-Hastings}
\label{sec:mh_mcmc}
MCMC is used to sample from high dimensional probability distributions. Consider a high-dimensional state space $X$, and a probability distribution function $p_X$ defined over it. Our goal is to generate samples $x$ from this distribution, denoted as $x \sim p_X$.
One such MCMC sampling algorithm is the Metropolis-Hastings (MH)~\cite{radfordneil}. MH is implemented by constructing two distributions:
\begin{itemize}
    \item \textbf{Proposal distribution:} Denoted as $q(x_i\rightarrow x_j)$, the proposal distribution $q$  specifies the probability of transitioning from each state in the state space $x_i$ to state $x_j$.
    \item \textbf{Acceptance distribution:} Denoted as $A(x_i\rightarrow x_j)$, the acceptance distribution determines whether a proposed transition is accepted. Specifically, it is defined as:
    \begin{equation}
        A(x_i\rightarrow x_j) = \min\left(1, \frac{p_X(x_j)q(x_j \rightarrow x_i)}{p_X(x_i)q(x_i \rightarrow x_j)}\right)
    \end{equation}
    $A(x_i \rightarrow x_j)$ represents the probability of accepting a move from $x_i$ to $x_j$, ensuring that the sampling process converges to the target distribution $p_X$.
\end{itemize}
The algorithm then proceeds as follows: Starting from a randomly initialized state $x_0$, MH generates a sequence of states $x_0\rightarrow x_1 \rightarrow ... \rightarrow x_T$.
At each step, a new state $x_j$ is sampled from current state $x_i$ using the proposal distribution $q(x_i \rightarrow x_j)$. This proposal is then accepted with probability $A(x_i \rightarrow x_j)$, resulting in the next state being $x_j$, or rejected, resulting in the next state remaining $x_i$. 
If the proposal distribution $q$ ensures that the state transitions are \emph{ergodic}, MH is guaranteed to asymptotically converge to a sample drawn from the distribution $p_X$. 

\subsubsection{3DGS Training with SGLD}
\label{sec:sgld_mcmc}
MH algorithm can be used to draw a sample $G$ from distribution $p$ in Eq.~\ref{eq:mcmcobjective} to obtain 3D Gaussian parameters. Starting from an random initial state $G_0$, MH generates a sequence of transitions $G_0\rightarrow G_1 \rightarrow G_2 .. G_t .. \rightarrow G_T$. The proposal distribution for transition from state $G_t$ to $G_{t+1}$ in the sequence is given by: 
\begin{equation}
    q(G_{t} \rightarrow G_{t+1}) = \mathcal{N}\left( G_t + \frac{\epsilon}{2}\nabla_G \log(p(G_t)), \epsilon I \right)
\end{equation}
where $\mathcal{N}$ is the standard normal distribution, $\epsilon$ is a step size parameter, $\nabla_G \log(p(G_t ))$ is the gradient of the log-likelihood of the parameters given the ground truth images, and $I$ is the identity matrix.
In standard MH, an acceptance distribution $A(G_t \rightarrow G_{t+1})$ would be used to determine whether to accept the proposed transition:
\begin{equation}
A(G_t \rightarrow G_{t+1}) = \min\left(1, \frac{p(G_{t+1})q(G_{t+1}\rightarrow G_{t})}{p(G_{t})q(G_{t}\rightarrow G_{t+1})} \right)    
\end{equation}

\noindent However, in the Stochastic Gradient Langevin Dynamics (SGLD) approximation, this explicit acceptance step is omitted. 
Specifically, the SGLD update rule is given by:
\begin{equation}
G_{t+1} = G_t + \frac{\epsilon}{2}\nabla_G \log(p(G_t)) + \sqrt{\epsilon}\eta    
\end{equation}
where $\eta \sim \mathcal{N}(0, I)$. 
\section{Related Work}
\label{sec:related_work}

\noindent \textbf{Compressing Scenes Represented by 3DGS.}
A range of prior works~\cite{lightgs, c3dgs, compactgs, pup3dgs, lp3dgs, rdgaussian, reducedgs, minisplatting, gaussianimage, hac, hac++, radsplat} aim to compress trained 3DGS models in order to reduce the storage required or to accelerate rendering speeds. To do this, these works propose several techniques such as pruning redundant gaussians, quantization, reducing bit-widths of less sensitive parameters, and cutting down higher order spherical harmonics coefficients to reduce the storage footprint needed to represent the scene. Other prior works~\cite{compact3d, lightgs, c3dgs} also propose using codebooks to compress the model memory footprint. 
However, these approaches compress the 3D model representation \emph{post-training}. In this work, we aim to enable both,  efficient training and rendering, by reducing the memory usage during the training process itself. To this end, we propose a novel codebook-based compression mechanism that significantly reduces memory utilization and model size during training. 



\noindent \textbf{Accelerating 3DGS Training.}
Prior works propose a range of techniques to accelerate 3DGS training.
SpeedySplat~\cite{speedysplat}, TurboGS~\cite{turbogs}, and TamingGS~\cite{taminggs} propose densification control heuristic strategies to reduce the number of gaussians and thus accelerate both training and rendering. These approaches reduce the memory consumption during training, however they incur some loss in representation quality. We quantitively compare against these works in Section~\ref{sec:results}, and we demonstrate that our approach achieves higher or similar compression rates without sacrificing representation quality.\footnote{We could not compare against TurboGS~\cite{turbogs} due to lack of availability of open-source code, but would incur the same quality challenges due to the reduced number of Gaussians for representation.}
Additionally, these approaches require careful tuning of the heuristic hyperparameters that may not generalize well across 3D scenes. 
EAGLES~\cite{eagles} and Scaffold-GS~\cite{scaffoldgs} aim to accelerate training speeds by using a compact representation that leverages neural networks to store 3D gaussian parameters more efficiently. We compare against EAGLES and Scaffold-GS in Section~\ref{sec:results} and demonstrate that it cannot achieve the same representation quality as state-of-art approaches. 


\noindent Recent works such as 3DGS-LM~\cite{3dgslm} and 3DGS$^2$~\cite{3dgs2} propose using second order optimization techniques that require fewer iterations for convergence, thus accelerating the training process. 
These approaches are orthogonal to our work and we note that 3DGS-LM only reports small speedups over the baseline~\cite{kerbl20233d}.
Other works~\cite{distwar, taminggs, flashgs, rtgs} propose low-level optimizations to the renderer's GPU code implementation or hardware support to speed up 3D gaussian splatting training~\cite{arc, gscore}. These training acceleration approaches are orthogonal to our work.

\section{Method}
\label{sec:method}


Our goal in this work is to enable training \emph{compressed representations} of 3D Gaussians for high-speed and memory efficient 3D reconstruction. 
To this end, we leverage codebook representations that can significantly reduce the amount of memory required to store 3DGS parameters. While codebooks have been demonstrated to be highly effective at compressing fully trained 3DGS models~\cite{c3dgs, compact3d, lightgs}, directly \emph{training} using codebook representations is an unsolved challenge. In this work, we solve the problem of learning on codebook-compressed representations by using MCMC sampling to sample over a posterior distribution of these compressed representations. 


\noindent This requires (1) defining a state space for codebook-compressed 3DGS parameters; 
(2) formulating a posterior distribution over the codebook-based state representation whose samples accurately reconstruct the 3D scene; and (3) defining state transitions and deriving corresponding proposal and acceptance probabilities for MCMC sampling.
We now describe these elements of our approach.

\subsection{Compressed Model State Space}
\label{sec:model_state}

\begin{figure}[!htb]
    \vspace{-0.4cm}
    \centering
    \includegraphics[width=\linewidth]{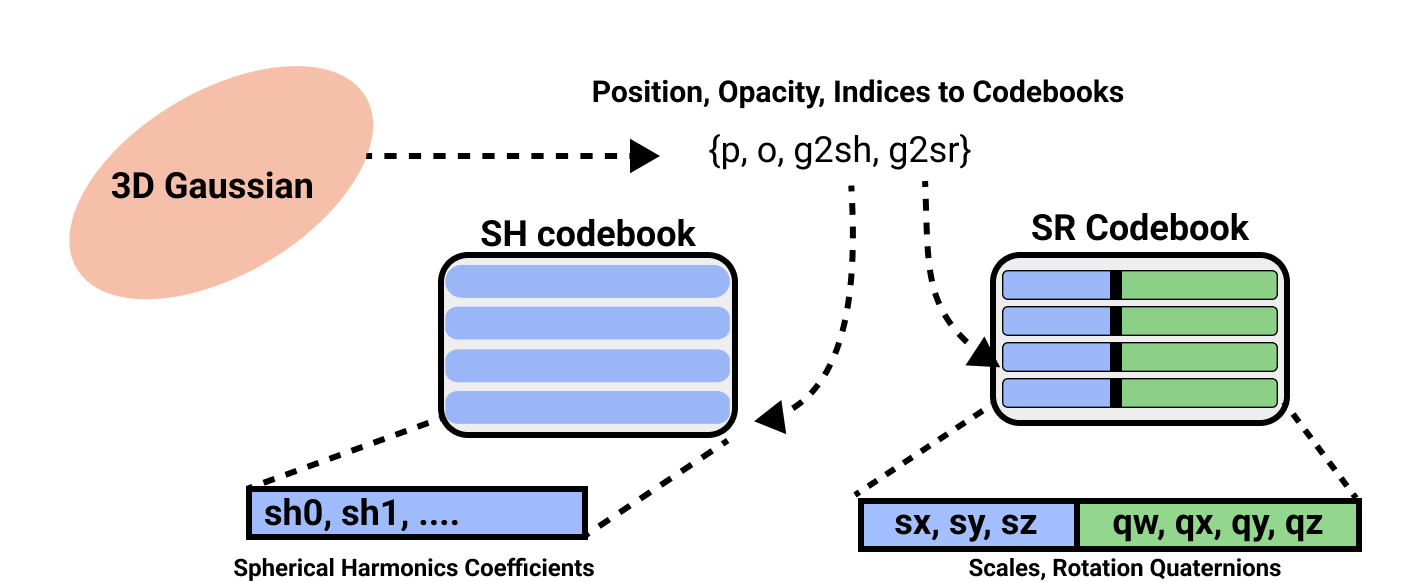}
    \caption{\textbf{Codebook layout used for \X.} \X allocates two codebooks: (1) SH stores spherical harmonics coefficientes of RGB colors, and (2) SR stores scaling + quaternion parameters concatenated together as vectors. Each Gaussian has g2sh, g2sr to index the two codebooks.}
    \label{fig:model_state_representation}
    \vspace{-0.3cm}
\end{figure}

We use a codebook representation similar to C3DGS~\cite{c3dgs}, but note that \X can also use a different codebook implementation (see Section~\ref{sec:background_codebooks}).
We define the state of the codebook-compressed 3DGS model as: $S= \{G, C\}$:
    \begin{equation}
        G = \{g_1, g_2, g_3, ... g_N\} \qquad C = \{SH, SR\}
    \end{equation}
In this expression, $C$ comprises the 3DGS codebooks: $SH$ and $SR$. The $SH$ codebook stores vectors of spherical harmonics coefficients used to derive the view dependent color. The $SR$ codebook contains vectors, where each vector stores the scaling and quaternion parameters for each Gaussian. $G$ corresponds to the set of 3D Gaussians, where $g_i$ corresponds to a Gaussian with the following parameters: 
    \begin{equation}
        g_i = \{\mathbf{p_i}, o_i, g2sr_i, g2sh_i\}
    \end{equation}
Here, $\boldsymbol{p_i}$ is the position of the mean, $o_i$ the opacity. $g2sh$ and $g2sr$ are integers that are pointers to vectors in the SH and SR codebook respectively.
Thus, a single 3D Gaussian in the compressed 3DGS model $G$ is represented as $s_i = \{g_i, C\}$. The state space representation described above is summarized in Fig.~\ref{fig:model_state_representation}. 

\subsection{Formulating The Posterior Distribution}
\label{sec:training_objective}
We define a posterior probability distribution over the  \X model state space as follows:
\begin{equation}
\label{eq:mcgprob}
p\left(G, C\right) \propto \mathrm{exp}\left(  -\mathcal{L}\left[ I_{\mathcal{P}}(G, C), I^{gt}_\mathcal{P} \right] \right)
\end{equation}
$I_\mathcal{P}(G, C)$ represents the image rendered from camera pose $\mathcal{P}$ using the \X codebook-based Gaussian model state $(G, C)$, and $I^{gt}_{\mathcal{P}}$ is the corresponding ground truth images. We define the reconstruction loss function $\mathcal{L}$ as:
\begin{equation}
    \label{eq:loss}
    \mathcal{L} = \mathcal{L}_{\text{recon}} + \lambda_{\text{sr}}|SR| +\lambda_{\text{sh}}|SH|
\end{equation}
$\mathcal{L}_{\text{recon}}$ is the reconstruction loss (defined in Eq.~\ref{eq:recon_loss}). The terms $|SR|$ and $|SH|$ represent the number of vectors in the SR and SH codebooks, respectively. $\lambda_{\text{sr}}$ and $\lambda_{\text{sh}}$ are hyperparameters used to control the size of these codebooks.

Bayesian inference over this probability function is performed by sampling the state $S = \{G, C\}$. Due to the high dimensionality of the distribution, we use MCMC sampling. However, unlike recent 3DGS-MCMC~\cite{3dgsmcmc} that uses SGLD for MCMC sampling (see Section~\ref{sec:sgld_mcmc}), SGLD cannot be applied to learn a discrete set of parameters such as the mapping/indexing between the Gaussians and codebook vectors. 
Instead we use the MH algorithm to sample from the posterior defined on codebook-compressed representations. To apply MH, we must define a proposal distribution and derive an acceptance distribution (see Section~\ref{sec:mh_mcmc}). The proposal distribution enables splitting and merging codebook vectors to transition between states with different mappings between Gaussians and codebook vectors. This allows MCMC to explore the joint space of continuous parameters. We now define the possible set of state transitions over codebook-based 3DGS representations, and define proposal distributions between these transitions.

\subsection{Model State Space Transitions}
\label{sec:state_transitions}
To sample from the probability distribution in Eq.~\ref{eq:mcgprob} using MH, we need to define a proposal and acceptance distribution functions (see Section~\ref{sec:mh_mcmc}). To do this, we first define a set of valid state transitions: (1) \textbf{parameter update}, (2) \textbf{merge} and (3) \textbf{split}.

\begin{figure}[!htb]
\vspace{-0.4cm}
  \centering
\includegraphics[width=0.8\linewidth]{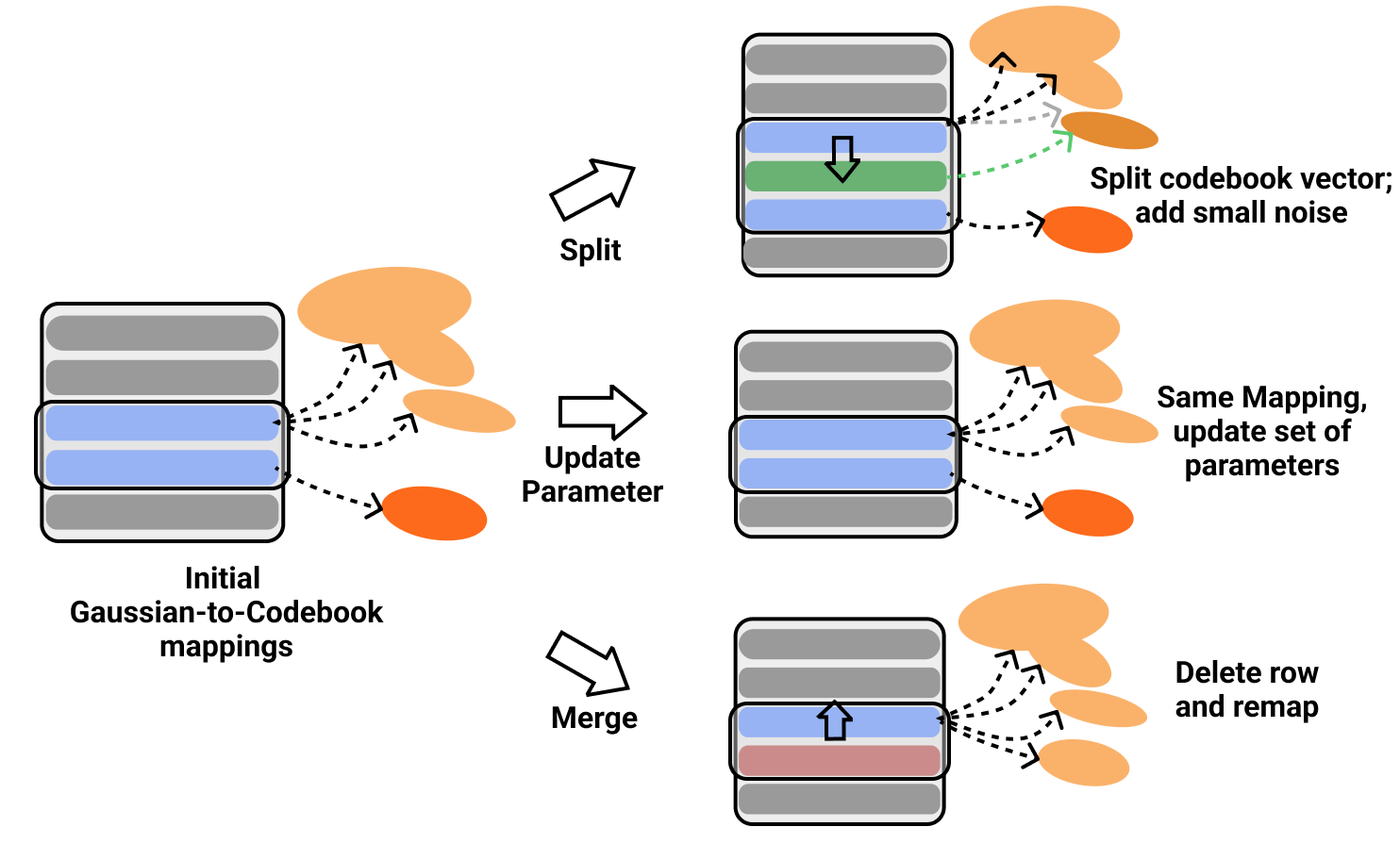}
\vspace{-0.2cm}
\caption{\X performs one of the 3 steps for each 3DG: (1) Updates Gaussian and codebook parameters, or (2) Splitting the 3DG's corresponding vector, or (3) Merge to the vector it previously split from.}
\label{fig:mcmc_state_transitions}
\vspace{-0.4cm}
\end{figure}

\begin{figure}[!htb]
    \vspace{-0.5cm}
  \begin{subfigure}{0.5\textwidth}
  \centering
    \includegraphics[width=0.9\linewidth]{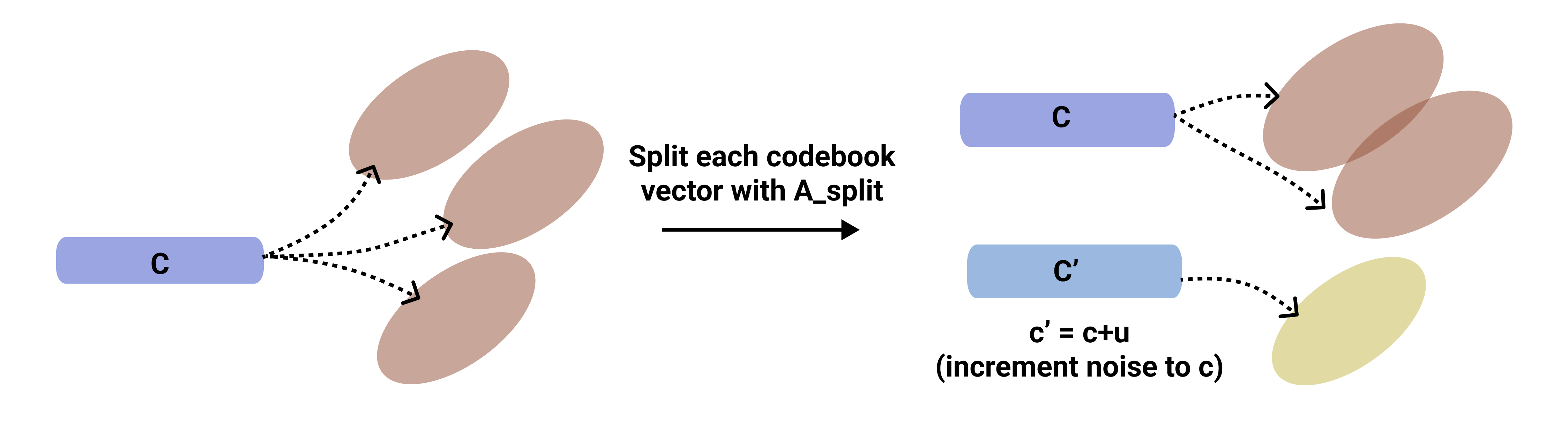}
    \caption{Split transition: a codebook vector referred to by two or more 3D Gaussian parameter is split into two vectors.}
    \label{fig:split}
  \end{subfigure}
  \hfill
  \begin{subfigure}{0.5\textwidth}
  \centering
    \includegraphics[width=0.9\linewidth]{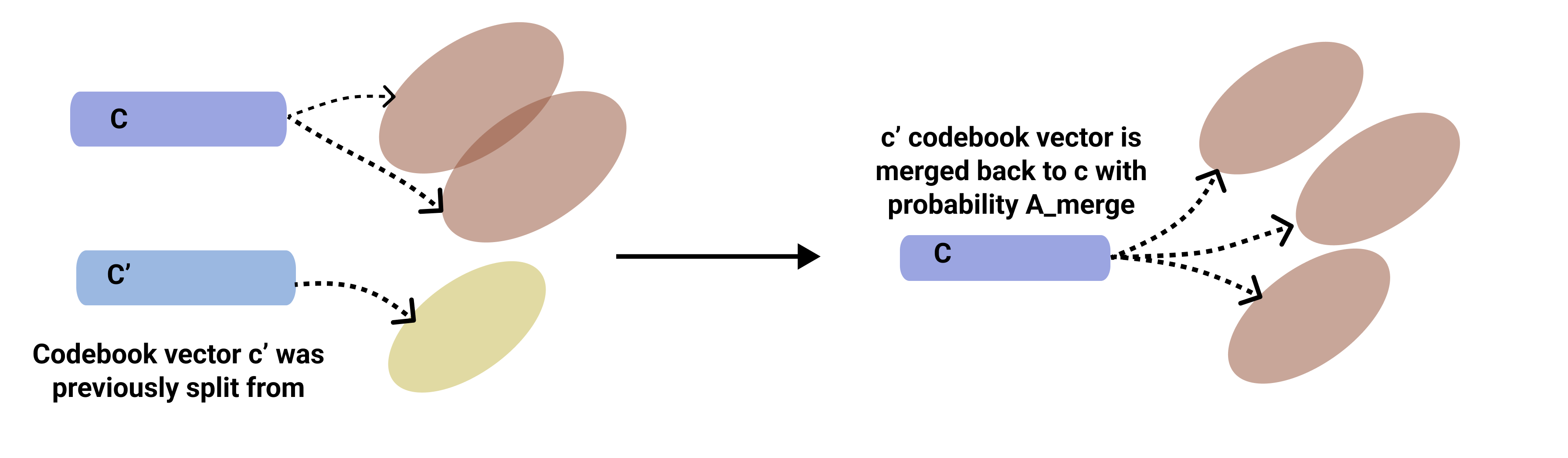}
    \caption{Merge transition: two codebook vectors referred to by different 3D Gaussian parameters gets merged into one.}
    \label{fig:merge}
  \end{subfigure}
  \caption{Split and Merge transitions.}
  \vspace{-0.2cm}  
  \label{fig:state_transitions}
  \vspace{-0.1cm}  
\end{figure}

\begin{itemize}
    \item \textbf{Parameter Update:} This involves changing the values of the parameters of the $SH,SR$ codebooks as well as 3D Gaussian parameters in $g_i$. The parameters are determined by the proposal distribution $q_{update}$. The mappings of the Gaussians to codebook vectors remains unchanged.

    \item \textbf{Split:} A Gaussian mapped to a codebook vector is split and allocated a new codebook vector, as shown in Fig.~\ref{fig:split}. The 3D Gaussian  mapped to codebook vector $c$ is allocated a new codebook vector $c'$. The value of $c'$ is generated by the proposal distribution $q_{split}$. 
    
    \item \textbf{Merge:} A Gaussian mapping to a codebook vector $c'$ can be merged with a codebook vector $c$ from which it was previously split from, as shown in Fig.~\ref{fig:merge}. The parameters in the codebook vector are unchanged. 
\end{itemize}

\noindent We now derive the proposal and acceptance distributions.

\subsection{Proposal and Acceptance Distributions}

We define the proposal distribution for MH sampling transition from state as: 
\begin{align}
q(S\rightarrow S')  = 0.98 q_{\text{update}}  + 0.01 q_{\text{split}} + 0.01 q_{\text{merge}}    
\end{align}

Where $q_{\text{update}}$ is the parameter update transition, $q_{\text{split}}$ is the split transition and $q_{\text{merge}}$ is the merge transition. In other words, at each step, we randomly choose the parameter update step $98\%$ of the times, and the split, merge transitions $1\%$ of the time. Now, we define the proposal distributions for each type of transition, and derive the acceptance distributions as in Section~\ref{sec:mh_mcmc}. Please refer to Section A.1
in the Appendix for a detailed derivation.


\noindent \textbf{Parameter update step} The proposal distribution for a parameter update, $q_{\text{update}}$, is identical to the SGLD parameter update, as described in Section~\ref{sec:sgld_mcmc}. Thus we express the parameter update state transition using $q_{\text{update}}$ as follows:
\begin{equation}
q_{\text{update}} (S_{\text{diff}}) = \mathcal{N}\left(\boldsymbol{S}_{\text{diff}} + \frac{\epsilon_p}{2}\nabla_{\boldsymbol{S}_{\text{diff}}} log (p (S)), \epsilon_p I\right)    
\end{equation}
Where $\epsilon_p$ is a small hyperparameters. $S_{\text{diff}}$ is the differentiable set of parameters (all parameters except the indices to the codebooks).
The update step is equivalent to the SGLD update step with acceptance $A=1$.

\noindent \textbf{Split Step:} 
During the split transition, for each Gaussian, we allocate a new codebook vector and remap. A Gaussian mapped to codebook vector $c$ is remapped to a new codebook vector entry row $c''$. The original codebook vector becomes $c'$. 
\begin{equation}
c' = c \qquad c'' = c + u    
\end{equation}
\begin{equation}
q_{\text{split}}(u) = \mathcal{N}(u|\mu=0, \sigma=\epsilon_{\text{split}}I)
\end{equation}
Where $\epsilon_{\text{split}}$ is a hyperparameter.

\noindent \textbf{Merge Step:} Two codebook vectors $c, c'$ can be merged into one codebook vector of value $c$. 
Two rows to be merged are selected with a transition distribution defined by:
\begin{equation}
q_{\text{merge}}(S\rightarrow S_{\text{merge}}) = \mathcal{N}(c-c'| \mu=0,\sigma=\epsilon_{\text{merge}} I)
\end{equation}
Where $\epsilon_{\text{merge}}$ is a hyperparameter.

\noindent \textbf{Acceptance:} Acceptance probabilities of the split and merge step can be derived as:
\begin{align}
A(S \rightarrow S_{\text{split}}) = \min \left(1, e^{-\lambda_{\text{SH}}}\frac{1}{q_{sm}(c'-c'')}\right)
\end{align}
\begin{align}
A(S \rightarrow S_{\text{merge}}) = \min\left(1,
e^{\lambda_{\text{SH}}}q_{sm}(c-c')\right)
\end{align}
Where $q_{sm}$ is given by:
\begin{equation}
    q_{sm}(u) = \exp\left(\frac{-u^2}{2}\left(\frac{1}{\epsilon^2_{split}} - \frac{1}{\epsilon^2_{merge}}\right) \right)
\end{equation}

\begin{table*}
\centering
\resizebox{0.8\textwidth}{!}{%
\begin{tabular}{l|cccc|cccc|cccc}
\hline
\cellcolor[HTML]{FFFFFF}{\color[HTML]{000000} } & \multicolumn{4}{c|}{\textbf{MipNerf360}} & \multicolumn{4}{c|}{\textbf{Deep Blending}} & \multicolumn{4}{c}{\textbf{Tanks and Temples}} \\ \hline
 & \textit{\textbf{PSNR}} & \textbf{SSIM} & \textbf{LPIPS} & \textbf{Peak Mem} & \textit{\textbf{PSNR}} & \textbf{SSIM} & \textbf{LPIPS} & \textbf{Peak Mem} & \textit{\textbf{PSNR}} & \textbf{SSIM} & \textbf{LPIPS} & \textbf{Peak Mem} \\ \hline
\textit{3DGS*} & 29.09 & 0.867 & 0.183 & 2089.363 & 30.10 & 0.909 & \cellcolor{yellow!25}0.241 & 817.159 & 22.03 & 0.821 & 0.197 & 901.405 \\
\textit{EAGLES*} & 28.70 & 0.867 & 0.194 & \cellcolor{orange!25}159.473 & 30.38 & 0.913 & 0.251 & \cellcolor{yellow!25}159.473 & 22.34 & 0.798 & 0.237 & \cellcolor{red!25}57.962 \\
\textit{SpeedySplat*} & \textit{28.33} & 0.846 & 0.241 & - & 30.02 & 0.907 & 0.269 & - & 21.68 & 0.773 & 0.289 & - \\
\textit{Reduced-GS*} & 29.03 & 0.870 & 0.184 & 1432.407 & 29.96 & 0.906 & 0.243 & 572.136 & 23.72 & 0.846 & \cellcolor{yellow!25}0.176 & 636.500 \\
\textit{Taming-GS*} & 29.39 & 0.863 & 0.198 & 477.598 & 30.11 & 0.910 & 0.251 & 325.629 & 22.58 & 0.830 & 0.191 & 1189.745 \\
\textit{Scaffold-GS*} & 29.22 & 0.869 & 0.190 & 304.827 & 30.89 & 0.913 & 0.244 & \cellcolor{red!25}95.107 & 22.54 & 0.829 & 0.190 & \cellcolor{yellow!25}192.211 \\ \hline
\textit{MCMC-2M} & \cellcolor{yellow!25}\textit{30.71} & \cellcolor{yellow!25}0.908 & \cellcolor{yellow!25}0.161 & 473.000 & \cellcolor{yellow!25}33.99 & \cellcolor{yellow!25}0.929 & 0.243 & 473.000 & 24.32 & \cellcolor{yellow!25}0.864 & 0.182 & 473.000 \\
\textit{Ours-2M} & 30.06 & 0.899 & 0.175 & \cellcolor{red!25}130.038 & 33.73 & 0.925 & 0.252 & \cellcolor{orange!25}120.672 & \cellcolor{yellow!25}24.35 & 0.861 & 0.187 & \cellcolor{orange!25}127.968 \\ \hline
\textit{MCMC-5M} & \cellcolor{red!25}\textit{32.72} & \cellcolor{red!25}0.943 & \cellcolor{red!25}0.128 & 1182.000 & \cellcolor{red!25}\textit{34.54} & \cellcolor{red!25}0.939 & \cellcolor{red!25}0.209 & 1182.000 & \cellcolor{red!25}26.63 & \cellcolor{red!25}0.911 & \cellcolor{red!25}0.112 & 1182.000 \\
\textit{Ours-5M} & \cellcolor{orange!25}31.01 & \cellcolor{orange!25}0.919 & \cellcolor{orange!25}0.146 & \cellcolor{yellow!25}275.594 & \cellcolor{orange!25}34.33 & \cellcolor{orange!25}0.932 & \cellcolor{orange!25}0.232 & 444.259 & \cellcolor{orange!25}\textit{26.18} & \cellcolor{orange!25}0.898 & \cellcolor{orange!25}0.127 & 338.000 \\
\hline
\end{tabular}%
}
\vspace{-0.2cm}
\caption{Averaged PSNR, SSIM and LPIPS and peak model memory during training (SfM initialization for * approaches)}
\label{tab:training_quality}
\vspace{-0.5cm}
\end{table*}


\section{Results}
\label{sec:results}

\vspace{-0.2cm}

\subsection{Experiment Setup}
\label{sec:experiment_setup}

\vspace{-0.2cm}

\textbf{Evaluation Platform, Hyperparameter Configuration.} We measure training and rendering speeds of \X on a system with a Core i7-13700K CPU and an NVIDIA RTX-4090 GPU. For our evaluation, we set $\lambda_{SH}$ to $2.3$, $\lambda_{SR}=3$. $\epsilon_{split}$ and $\epsilon_{merge}$ are set to $0.1, 0.05$ for both SH and SR codebooks. At each training step, we choose to perform the parameter update step with a probability of $98\%$, and the split and merge transitions with a $1\%$ probability. We initialize training with 100000 Gaussians with random parameters and grow the number of Gaussians by $5\%$ every $100$ training iterations. We evaluate two versions of \X, \X-2M and \X-5M in which we cap the Gaussians count at 2 million and 5 million respectively.

\noindent \textbf{Dataset and Metrics.} We evaluate \X on 3D reconstruction tasks using multi-view images.
We measure the PSNR, LPIPS and SSIM on evaluation views of 3D reconstruction tasks. 
We consider the following datasets that contain multiview images of real world and synthetic 3D scenes: MipNerf360~\cite{m360} (counter, stump, kitchen, bicycle, bonsai, room and garden scenes), Blender~\cite{nerf} (chair, drums, ficus, hotdog, lego, materials, mic and ship scenes), Deep Blending's Playroom scene and Tanks and Temples~\cite{tandt} (Truck and Train scenes) datasets. 

\noindent \textbf{Prior Work Comparisons.} We then compare the reconstruction quality with prior works.  We compare \X approach against 3DGS~\cite{kerbl20233d}, MCMC-2M~\cite{3dgsmcmc}, and MCMC-5M (two configurations with 2 million and 5 million Gaussians respectively); 3DGS-MCMC~\cite{3dgsmcmc} achieves state-of-art representation quality. Both MCMC-5M and \X-5M incur higher memory cost than their respective 2M configurations but provide higher representation qualities.
We also compare with other prior works that enable faster and more memory efficient training by pruning the number of Gaussians (Section~\ref{sec:related_work}): EAGLES~\cite{eagles}, SpeedySplat~\cite{speedysplat}, Reduced-GS\cite{reducedgs}, TamingGS~\cite{taminggs}, Scaffold-GS~\cite{scaffoldgs}. For prior work, we initialize the positions and RGB colors of the Gaussians using the corresponding Structure-from-Motion point cloud.

\begin{figure*}
\centering
\begin{subfigure}{0.33\linewidth}
\centering\captionsetup{width=0.96\linewidth}%
    \includegraphics[width=0.9\linewidth]{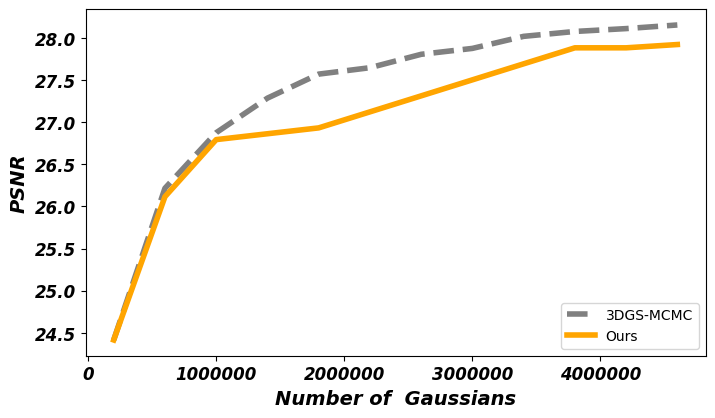}
    \vspace{-0.1cm}
    \caption{\#Gaussians vs. PSNR}
    \label{fig:psnr_num3dg}
\end{subfigure}
\begin{subfigure}{0.33\linewidth}
\centering\captionsetup{width=0.96\linewidth}%
    \includegraphics[width=0.9\linewidth]{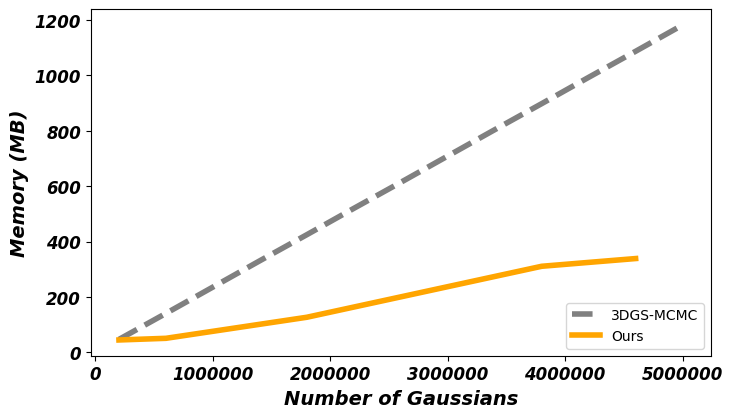}
    \vspace{-0.1cm}
    \caption{\#Gaussians vs. Model Memory}
    \label{fig:mem_num3dg}
\end{subfigure}
\begin{subfigure}{0.33\linewidth}
\centering\captionsetup{width=0.96\linewidth}%
    \includegraphics[width=0.9\linewidth]{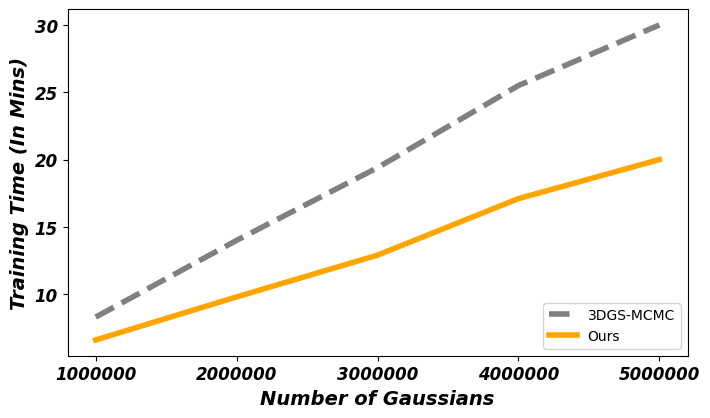}
    \vspace{-0.1cm}
    \caption{\#Gaussians vs. Training Time}
    \label{fig:train_num3dg}
\end{subfigure}
\vspace{-0.4cm}
\caption{The impact of Gaussian count for 3DGS-MCMC and \X on training quality (PSNR), model memory, and training time. }
\vspace{-0.4cm}
\end{figure*}

\subsection{Representation Quality \& Peak Model Memory}
\label{sec:quality_metric_results}

Table~\ref{tab:training_quality} shows the average PSNR, SSIM, LPIPS and the peak model memory during training (in MB) achieved by \X on multiview 3D scene reconstruction tasks. PSNR, SSIM and LPIPS measured for individual scenes are presented in Section B
of Appendix along with qualitative comparisons of generated images.  

\noindent First, we observe that \X is able to achieve similar representation quality as that of a state-of-art approach 3DGS-MCMC~\cite{3dgsmcmc}: less than 0.3 PSNR on average with 2 million Gaussians, and less than 0.8 PSNR on average with 5 million Gaussians. \X-2M achieves PSNR equivalent to MCMC-2M on the Deep Blending and Tanks and Temples datasets, and incurs a 0.65 PSNR drop on average on the MipNerf360 dataset. \X-5M also achieves equivalent PSNR on the DeepBlending incurs a sharper PSNR drop on MipNerf360 (1.5 PSNR).  At the same time, \X requires significantly less peak model memory ($3.78\times$  and $3.5\times$ on average respectively). 
Second, \X-5M outperforms MCMC-2M on all quality metrics (PSNR, SSIM and LPIPS) while having smaller model memory size ($1.39\times$ on average). 
Third, compared to the prior work that aims to achieve memory efficient training, \X-2M achieves significantly higher representation quality while having a similar peak memory usage compared to \todo{}
EAGLES, Reduced-GS, Taming-GS and Scaffold-GS.
Finally, compared to 3DGS, \X-2M is able to significantly reduce the peak model memory during training, by $9.99\times$ on average.
We conclude that \X enables significantly reduced peak memory during training while retaining close to state-of-art quality. Thus, for any memory capacity constraint (i.e, peak memory utilization), \X achieves the highest representation quality compared to prior works.


\noindent \textbf{Random point cloud initialization.} We note that \X-2M and \X-5M (and MCMC-2M, 5M) are able to reconstruct the scene initialized on a random point cloud). The results presented in this section for these configurations are with randomly initialized Gaussian parameters.
EAGLES, Taming-GS, SpeedySplat, Reduced-GS require the SfM point cloud to initialize the Gaussian parameters and the results in this section were collected using SfM. 
A fairer comparison between these methods would assume a random initialization, but since certain scenes in the MipNerf360, Deep Blending, Tanks and Temples datasets could not be trained with random initialization for these approaches, the results in Table~\ref{tab:training_quality} include SfM initialization for all approaches except \X and MCMC. We were able to train scenes in the Blender dataset for all approaches (except Taming-GS) with random initialization, and we present these results in Table~\ref{tab:training_quality_random_init}. 


\noindent Table~\ref{tab:training_quality_random_init} shows the average quality metrics and peak memory during training of \X-2M for synthetic scenes in the Blender dataset. \X-2M achieves a significantly higher reconstruction quality compared to prior works. Although the peak model memory during training of \X-2M is higher in comparison, the peak memory is in the same range as that of real world scenes in Table~\ref{tab:training_quality}. We do not report results from MCMC here because MCMC-2M fails to reconstruct all synthetic scenes when setting the Gaussians count to 2M, due to instability in the training process.

\begin{table}[!htb]
\vspace{-0.4cm}
\centering
\resizebox{0.7\columnwidth}{!}{%
\begin{tabular}{c|cccc}
\hline
 & \textit{\textbf{PSNR}} & \textbf{SSIM} & \textbf{LPIPS} & \textbf{Peak Mem} \\ \hline
3DGS & 31.07 & 0.959 & 0.051 & 83.00 \\
\textit{EAGLES} & 32.55 & 0.964 & 0.039 & \cellcolor{red!25}5.17 \\
\textit{SpeedySplat} & 32.43 & 0.960 & 0.050 & \cellcolor{yellow!25}47.89 \\
Reduced-GS & \cellcolor{orange!25}33.80 & \cellcolor{orange!25}0.970 & \cellcolor{orange!25}0.030 & 49.72 \\
\textit{Scaffold-GS} & \cellcolor{yellow!25}33.46 & \cellcolor{yellow!25}0.967 & \cellcolor{yellow!25}0.034 & \cellcolor{orange!25}31.66 \\
\textit{Ours-2M} & \cellcolor{red!25}\textit{36.840} & \cellcolor{red!25}0.984 & \cellcolor{red!25}0.023 & 257.87 \\ \hline
\end{tabular}%
}
\caption{PSNR, SSIM, LPIPS for different workloads on the Blender dataset (GS parameters initialized randomly)}
\label{tab:training_quality_random_init}
\vspace{-0.4cm}
\end{table}




\subsection{Training and Rendering Speeds}
\label{sec:training_efficiency}

Table~\ref{tab:training_efficiency} shows the average training speed in terms of training iterations per second for \X. We observe that, when compared to MCMC-2M, \X-2M is able to speed up training and rendering by $1.36\times$ and $1.88\times$ on average. We also compare against Taming3DGS, the state-of-art for efficient training, in Section E
of the Appendix. These works achieve higher training and rendering speeds. However, the representation quality is significantly lower.


\begin{table}[!htb]
\centering
\resizebox{0.9\columnwidth}{!}{%
\small
\begin{tabular}{l|cc|cc|cc}
\hline
\multicolumn{1}{c|}{\textit{}} & \multicolumn{2}{c|}{\textbf{MipNerf360}}  & \multicolumn{2}{c|}{\textbf{Deep Blending}} & \multicolumn{2}{c}{\textbf{T \& T}} \\ \hline
\multicolumn{1}{c|}{} & \textbf{Train It/s} & \textbf{FPS} & \textbf{Train It/s} & \textbf{FPS} & \textbf{Train It/s} & \textbf{FPS}  \\ \hline
MCMC & 23.65 & 102 & 31.66 & 161  & 30.34 & 133 \\
Ours & 32.02 & 216 & 46.33 & 319  & 40.75 & 207 \\
\hline
MCMC-5M & 11.71 & 65 & 14.06 & 114  & 12.95 & 80 \\
Ours-5M & 21.1 & 87 & 19.87 & 141  & 19.18 & 114 \\
\hline
\end{tabular}%
}
\vspace{-0.2cm}
\caption{Training and rendering speeds}
\label{tab:training_efficiency}
\vspace{-0.3cm}
\end{table}

\subsection{Ablation Study}
\label{sec:ablations}

\noindent \textbf{Quality vs. Gaussians Count.} Figure~\ref{fig:psnr_num3dg} depicts the PSNR achieved by \X and 3DGS-MCMC on varying the number of Gaussians used to represent the Truck scene of the T\&T dataset. We observe that as the Gaussian count increases, the PSNR achieved by \X incurs a small degradation in quality compared to 3DGS-MCMC. At 5 million Gaussians, the difference in PSNR is about $0.46$ on average. Similar to 3DGS-MCMC, \X achieves higher representation quality with a greater Gaussian count.





\noindent \textbf{Model memory vs. Number of Gaussians.} Fig.~\ref{fig:mem_num3dg} depicts how the post-training memory footprint of the model varies with the Gaussians count for the Tanks and Temples dataset. This memory footprint increases linearly with the Gaussian count for 3DGS-MCMC. However, \X is able to represent large Gaussian counts without significantly increasing the memory footprint and requires a lot less memory than 3DGS-MCMC ($3.49\times$ on average). 



\noindent \textbf{Training speed vs. Number of Gaussians.}
Fig.~\ref{fig:train_num3dg} depicts the wall clock time needed to train the playroom scene of the Deep Blending dataset for 30000 iterations. With 3DGS-MCMC, the training time increases linearly with the gaussian count, however, \X enable more efficient and scalable training at larger gaussian counts. 






\section{Conclusion}
\label{sec:conclusion}

We introduce \X, a novel method for learning codebook-compressed representations of 3D Gaussian Splatting for high-speed memory-efficient scene reconstruction.
\X can effectively reduce the model's memory footprint during training while accelerating training and rendering. 
This makes \X useful for reconstructing complex scenes with 3DGS, which is challenging on resource-constrained platforms such as mobile devices and web browsers.
\X achieves the highest representation quality for any memory capacity constraint (i.e, peak memory utilization) compared to existing approaches. 
\X provides an extensible mathematical framework that can be applied to do compressed training on a range of point-based scene representation methods.


{
    \small
    \bibliographystyle{ieeenat_fullname}
    \bibliography{main}

\begin{thebibliography}{71}
\providecommand{\natexlab}[1]{#1}
\providecommand{\url}[1]{\texttt{#1}}
\expandafter\ifx\csname urlstyle\endcsname\relax
  \providecommand{\doi}[1]{doi: #1}\else
  \providecommand{\doi}{doi: \begingroup \urlstyle{rm}\Url}\fi

\bibitem[Abou-Chakra et~al.(2024)Abou-Chakra, Rana, Dayoub, and S{\"u}nderhauf]{physicallyembodied-gs}
Jad Abou-Chakra, Krishan Rana, Feras Dayoub, and Niko S{\"u}nderhauf.
\newblock Physically embodied gaussian splatting: A realtime correctable world model for robotics.
\newblock \emph{arXiv preprint arXiv:2406.10788}, 2024.

\bibitem[Barron et~al.(2022)Barron, Mildenhall, Verbin, Srinivasan, and Hedman]{m360}
Jonathan~T Barron, Ben Mildenhall, Dor Verbin, Pratul~P Srinivasan, and Peter Hedman.
\newblock Mip-nerf 360: Unbounded anti-aliased neural radiance fields.
\newblock In \emph{Proceedings of the IEEE/CVF conference on computer vision and pattern recognition}, pages 5470--5479, 2022.

\bibitem[Chen et~al.(2024{\natexlab{a}})Chen, Laina, and Vedaldi]{direct-dge}
Minghao Chen, Iro Laina, and Andrea Vedaldi.
\newblock Dge: Direct gaussian 3d editing by consistent multi-view editing.
\newblock In \emph{European Conference on Computer Vision}, pages 74--92. Springer, 2024{\natexlab{a}}.

\bibitem[Chen et~al.(2024{\natexlab{b}})Chen, Chen, Zhang, Wang, Yang, Wang, Cai, Yang, Liu, and Lin]{gaussianeditor}
Yiwen Chen, Zilong Chen, Chi Zhang, Feng Wang, Xiaofeng Yang, Yikai Wang, Zhongang Cai, Lei Yang, Huaping Liu, and Guosheng Lin.
\newblock Gaussianeditor: Swift and controllable 3d editing with gaussian splatting.
\newblock In \emph{Proceedings of the IEEE/CVF conference on computer vision and pattern recognition}, pages 21476--21485, 2024{\natexlab{b}}.

\bibitem[Chen et~al.(2024{\natexlab{c}})Chen, Wu, Lin, Harandi, and Cai]{hac}
Yihang Chen, Qianyi Wu, Weiyao Lin, Mehrtash Harandi, and Jianfei Cai.
\newblock Hac: Hash-grid assisted context for 3d gaussian splatting compression.
\newblock In \emph{European Conference on Computer Vision}, pages 422--438. Springer, 2024{\natexlab{c}}.

\bibitem[Chen et~al.(2025)Chen, Wu, Lin, Harandi, and Cai]{hac++}
Yihang Chen, Qianyi Wu, Weiyao Lin, Mehrtash Harandi, and Jianfei Cai.
\newblock Hac++: Towards 100x compression of 3d gaussian splatting.
\newblock \emph{arXiv preprint arXiv:2501.12255}, 2025.

\bibitem[Deng et~al.(2024)Deng, Chen, Zhang, Yang, Yuan, Liu, Wang, Wang, and Chen]{compactgs}
Tianchen Deng, Yaohui Chen, Leyan Zhang, Jianfei Yang, Shenghai Yuan, Jiuming Liu, Danwei Wang, Hesheng Wang, and Weidong Chen.
\newblock Compact 3d gaussian splatting for dense visual slam.
\newblock \emph{arXiv preprint arXiv:2403.11247}, 2024.

\bibitem[Durvasula et~al.(2023)Durvasula, Zhao, Chen, Liang, Sanjaya, and Vijaykumar]{distwar}
Sankeerth Durvasula, Adrian Zhao, Fan Chen, Ruofan Liang, Pawan~Kumar Sanjaya, and Nandita Vijaykumar.
\newblock Distwar: Fast differentiable rendering on raster-based rendering pipelines.
\newblock \emph{arXiv preprint arXiv:2401.05345}, 2023.

\bibitem[Durvasula et~al.(2025)Durvasula, Zhao, Chen, Liang, Sanjaya, Guan, Giannoula, and Vijaykumar]{arc}
Sankeerth Durvasula, Adrian Zhao, Fan Chen, Ruofan Liang, Pawan~Kumar Sanjaya, Yushi Guan, Christina Giannoula, and Nandita Vijaykumar.
\newblock Arc: Warp-level adaptive atomic reduction in gpus to accelerate differentiable rendering.
\newblock In \emph{Proceedings of the 30th ACM International Conference on Architectural Support for Programming Languages and Operating Systems, Volume 1}, pages 64--83, 2025.

\bibitem[Fan et~al.(2025)Fan, Wang, Wen, Zhu, Xu, Wang, et~al.]{lightgs}
Zhiwen Fan, Kevin Wang, Kairun Wen, Zehao Zhu, Dejia Xu, Zhangyang Wang, et~al.
\newblock Lightgaussian: Unbounded 3d gaussian compression with 15x reduction and 200+ fps.
\newblock \emph{Advances in neural information processing systems}, 37:\penalty0 140138--140158, 2025.

\bibitem[Fang and Wang(2024)]{minisplatting}
Guangchi Fang and Bing Wang.
\newblock Mini-splatting: Representing scenes with a constrained number of gaussians.
\newblock In \emph{European Conference on Computer Vision}, pages 165--181. Springer, 2024.

\bibitem[Feng et~al.(2024)Feng, Chen, Fu, Liao, Wang, Liu, Pei, Li, Zhang, and Dai]{flashgs}
Guofeng Feng, Siyan Chen, Rong Fu, Zimu Liao, Yi Wang, Tao Liu, Zhilin Pei, Hengjie Li, Xingcheng Zhang, and Bo Dai.
\newblock Flashgs: Efficient 3d gaussian splatting for large-scale and high-resolution rendering.
\newblock \emph{arXiv preprint arXiv:2408.07967}, 2024.

\bibitem[Gao et~al.(2024{\natexlab{a}})Gao, Xu, Cao, Mildenhall, Ma, Chen, Tang, and Neumann]{gaussianflow}
Quankai Gao, Qiangeng Xu, Zhe Cao, Ben Mildenhall, Wenchao Ma, Le Chen, Danhang Tang, and Ulrich Neumann.
\newblock Gaussianflow: Splatting gaussian dynamics for 4d content creation.
\newblock \emph{arXiv preprint arXiv:2403.12365}, 2024{\natexlab{a}}.

\bibitem[Gao et~al.(2024{\natexlab{b}})Gao, Holynski, Henzler, Brussee, Martin-Brualla, Srinivasan, Barron, and Poole]{cat3d}
Ruiqi Gao, Aleksander Holynski, Philipp Henzler, Arthur Brussee, Ricardo Martin-Brualla, Pratul Srinivasan, Jonathan~T Barron, and Ben Poole.
\newblock Cat3d: Create anything in 3d with multi-view diffusion models.
\newblock \emph{arXiv preprint arXiv:2405.10314}, 2024{\natexlab{b}}.

\bibitem[Girish et~al.(2024)Girish, Gupta, and Shrivastava]{eagles}
Sharath Girish, Kamal Gupta, and Abhinav Shrivastava.
\newblock Eagles: Efficient accelerated 3d gaussians with lightweight encodings.
\newblock In \emph{European Conference on Computer Vision}, pages 54--71. Springer, 2024.

\bibitem[Govindarajan et~al.(2025)Govindarajan, Rebain, Yi, and Tagliasacchi]{radiantfoam}
Shrisudhan Govindarajan, Daniel Rebain, Kwang~Moo Yi, and Andrea Tagliasacchi.
\newblock Radiant foam: Real-time differentiable ray tracing.
\newblock \emph{arXiv preprint arXiv:2502.01157}, 2025.

\bibitem[Hanson et~al.(2024{\natexlab{a}})Hanson, Tu, Lin, Singla, Zwicker, and Goldstein]{speedysplat}
Alex Hanson, Allen Tu, Geng Lin, Vasu Singla, Matthias Zwicker, and Tom Goldstein.
\newblock Speedy-splat: Fast 3d gaussian splatting with sparse pixels and sparse primitives.
\newblock \emph{arXiv preprint arXiv:2412.00578}, 2024{\natexlab{a}}.

\bibitem[Hanson et~al.(2024{\natexlab{b}})Hanson, Tu, Singla, Jayawardhana, Zwicker, and Goldstein]{pup3dgs}
Alex Hanson, Allen Tu, Vasu Singla, Mayuka Jayawardhana, Matthias Zwicker, and Tom Goldstein.
\newblock Pup 3d-gs: Principled uncertainty pruning for 3d gaussian splatting.
\newblock \emph{arXiv preprint arXiv:2406.10219}, 2024{\natexlab{b}}.

\bibitem[Hastings(1970)]{radfordneil}
W.~K. Hastings.
\newblock Monte carlo sampling methods using markov chains and their applications.
\newblock \emph{Biometrika}, 57\penalty0 (1):\penalty0 97--109, 1970.

\bibitem[H{\"o}llein et~al.(2024)H{\"o}llein, Bo{\v{z}}i{\v{c}}, Zollh{\"o}fer, and Nie{\ss}ner]{3dgslm}
Lukas H{\"o}llein, Alja{\v{z}} Bo{\v{z}}i{\v{c}}, Michael Zollh{\"o}fer, and Matthias Nie{\ss}ner.
\newblock 3dgs-lm: Faster gaussian-splatting optimization with levenberg-marquardt.
\newblock \emph{arXiv preprint arXiv:2409.12892}, 2024.

\bibitem[Hu et~al.(2024)Hu, Chen, Feng, Li, Yang, Bao, Zhang, and Cui]{cg-slam}
Jiarui Hu, Xianhao Chen, Boyin Feng, Guanglin Li, Liangjing Yang, Hujun Bao, Guofeng Zhang, and Zhaopeng Cui.
\newblock Cg-slam: Efficient dense rgb-d slam in a consistent uncertainty-aware 3d gaussian field.
\newblock In \emph{European Conference on Computer Vision}, pages 93--112. Springer, 2024.

\bibitem[Hu et~al.(2025)Hu, Liu, Chen, Feng, and Beerel]{splatmap}
Yue Hu, Rong Liu, Meida Chen, Andrew Feng, and Peter Beerel.
\newblock Splatmap: Online dense monocular slam with 3d gaussian splatting.
\newblock \emph{arXiv preprint arXiv:2501.07015}, 2025.

\bibitem[Jiang et~al.(2024)Jiang, Zheng, Lyu, Zhou, and Wang]{brightdreamer}
Lutao Jiang, Xu Zheng, Yuanhuiyi Lyu, Jiazhou Zhou, and Lin Wang.
\newblock Brightdreamer: Generic 3d gaussian generative framework for fast text-to-3d synthesis.
\newblock \emph{arXiv preprint arXiv:2403.11273}, 2024.

\bibitem[Jung et~al.(2023)Jung, Brasch, Song, Perez-Pellitero, Zhou, Li, Navab, and Busam]{pardy-human}
HyunJun Jung, Nikolas Brasch, Jifei Song, Eduardo Perez-Pellitero, Yiren Zhou, Zhihao Li, Nassir Navab, and Benjamin Busam.
\newblock Deformable 3d gaussian splatting for animatable human avatars.
\newblock \emph{arXiv preprint arXiv:2312.15059}, 2023.

\bibitem[Kerbl et~al.(2023)Kerbl, Kopanas, Leimk{\"u}hler, and Drettakis]{kerbl20233d}
Bernhard Kerbl, Georgios Kopanas, Thomas Leimk{\"u}hler, and George Drettakis.
\newblock 3d gaussian splatting for real-time radiance field rendering.
\newblock \emph{ACM Trans. Graph.}, 42\penalty0 (4):\penalty0 139--1, 2023.

\bibitem[Kheradmand et~al.(2024)Kheradmand, Rebain, Sharma, Sun, Tseng, Isack, Kar, Tagliasacchi, and Yi]{3dgsmcmc}
Shakiba Kheradmand, Daniel Rebain, Gopal Sharma, Weiwei Sun, Jeff Tseng, Hossam Isack, Abhishek Kar, Andrea Tagliasacchi, and Kwang~Moo Yi.
\newblock 3d gaussian splatting as markov chain monte carlo.
\newblock \emph{arXiv preprint arXiv:2404.09591}, 2024.

\bibitem[Knapitsch et~al.(2017)Knapitsch, Park, Zhou, and Koltun]{tandt}
Arno Knapitsch, Jaesik Park, Qian-Yi Zhou, and Vladlen Koltun.
\newblock Tanks and temples: Benchmarking large-scale scene reconstruction.
\newblock \emph{ACM Transactions on Graphics}, 36\penalty0 (4), 2017.

\bibitem[Kocabas et~al.(2024)Kocabas, Chang, Gabriel, Tuzel, and Ranjan]{humansplat-kocabas}
Muhammed Kocabas, Jen-Hao~Rick Chang, James Gabriel, Oncel Tuzel, and Anurag Ranjan.
\newblock Hugs: Human gaussian splats.
\newblock In \emph{Proceedings of the IEEE/CVF conference on computer vision and pattern recognition}, pages 505--515, 2024.

\bibitem[Lan et~al.(2025)Lan, Shao, Lu, Zhang, Jiang, and Yang]{3dgs2}
Lei Lan, Tianjia Shao, Zixuan Lu, Yu Zhang, Chenfanfu Jiang, and Yin Yang.
\newblock 3dgs$^2$: Near second-order converging 3d gaussian splatting.
\newblock \emph{arXiv preprint arXiv:2501.13975}, 2025.

\bibitem[Lee et~al.(2024{\natexlab{a}})Lee, Lee, Lee, Park, and Sim]{gscore}
Junseo Lee, Seokwon Lee, Jungi Lee, Junyong Park, and Jaewoong Sim.
\newblock Gscore: Efficient radiance field rendering via architectural support for 3d gaussian splatting.
\newblock In \emph{Proceedings of the 29th ACM International Conference on Architectural Support for Programming Languages and Operating Systems, Volume 3}, pages 497--511, 2024{\natexlab{a}}.

\bibitem[Lee et~al.(2024{\natexlab{b}})Lee, Rho, Sun, Ko, and Park]{compact3d}
Joo~Chan Lee, Daniel Rho, Xiangyu Sun, Jong~Hwan Ko, and Eunbyung Park.
\newblock Compact 3d gaussian representation for radiance field.
\newblock In \emph{Proceedings of the IEEE/CVF Conference on Computer Vision and Pattern Recognition}, pages 21719--21728, 2024{\natexlab{b}}.

\bibitem[Li et~al.(2024)Li, Li, Zhang, Zhang, Jia, Wang, Fan, Tseng, and Wang]{robogsim}
Xinhai Li, Jialin Li, Ziheng Zhang, Rui Zhang, Fan Jia, Tiancai Wang, Haoqiang Fan, Kuo-Kun Tseng, and Ruiping Wang.
\newblock Robogsim: A real2sim2real robotic gaussian splatting simulator.
\newblock \emph{arXiv preprint arXiv:2411.11839}, 2024.

\bibitem[Lin et~al.(2024)Lin, Feng, and Zhu]{rtgs}
Weikai Lin, Yu Feng, and Yuhao Zhu.
\newblock Rtgs: Enabling real-time gaussian splatting on mobile devices using efficiency-guided pruning and foveated rendering.
\newblock \emph{arXiv preprint arXiv:2407.00435}, 2024.

\bibitem[Liu et~al.(2024)Liu, Xu, Hu, Chen, and Feng]{atomgs}
Rong Liu, Rui Xu, Yue Hu, Meida Chen, and Andrew Feng.
\newblock Atomgs: Atomizing gaussian splatting for high-fidelity radiance field.
\newblock \emph{arXiv preprint arXiv:2405.12369}, 2024.

\bibitem[Liu et~al.(2025)Liu, Sun, Chen, Wang, and Feng]{dbsplatting}
Rong Liu, Dylan Sun, Meida Chen, Yue Wang, and Andrew Feng.
\newblock Deformable beta splatting.
\newblock \emph{arXiv preprint arXiv:2501.18630}, 2025.

\bibitem[Lu et~al.(2024{\natexlab{a}})Lu, Dhiman, Srinath, Arslan, Xing, Xiangli, Babu, and Sridhar]{turbogs}
Tao Lu, Ankit Dhiman, R Srinath, Emre Arslan, Angela Xing, Yuanbo Xiangli, R~Venkatesh Babu, and Srinath Sridhar.
\newblock Turbo-gs: Accelerating 3d gaussian fitting for high-quality radiance fields.
\newblock \emph{arXiv preprint arXiv:2412.13547}, 2024{\natexlab{a}}.

\bibitem[Lu et~al.(2024{\natexlab{b}})Lu, Yu, Xu, Xiangli, Wang, Lin, and Dai]{scaffoldgs}
Tao Lu, Mulin Yu, Linning Xu, Yuanbo Xiangli, Limin Wang, Dahua Lin, and Bo Dai.
\newblock Scaffold-gs: Structured 3d gaussians for view-adaptive rendering.
\newblock In \emph{Proceedings of the IEEE/CVF Conference on Computer Vision and Pattern Recognition}, pages 20654--20664, 2024{\natexlab{b}}.

\bibitem[Lu et~al.(2024{\natexlab{c}})Lu, Yu, Xu, Xiangli, Wang, Lin, and Dai]{scaffolds}
Tao Lu, Mulin Yu, Linning Xu, Yuanbo Xiangli, Limin Wang, Dahua Lin, and Bo Dai.
\newblock Scaffold-gs: Structured 3d gaussians for view-adaptive rendering.
\newblock In \emph{Proceedings of the IEEE/CVF Conference on Computer Vision and Pattern Recognition}, pages 20654--20664, 2024{\natexlab{c}}.

\bibitem[Mallick et~al.(2024)Mallick, Goel, Kerbl, Steinberger, Carrasco, and De~La~Torre]{taminggs}
Saswat~Subhajyoti Mallick, Rahul Goel, Bernhard Kerbl, Markus Steinberger, Francisco~Vicente Carrasco, and Fernando De~La~Torre.
\newblock Taming 3dgs: High-quality radiance fields with limited resources.
\newblock In \emph{SIGGRAPH Asia 2024 Conference Papers}, pages 1--11, 2024.

\bibitem[Mihajlovic et~al.(2024)Mihajlovic, Prokudin, Tang, Maier, Bogo, Tung, and Boyer]{splatfield}
Marko Mihajlovic, Sergey Prokudin, Siyu Tang, Robert Maier, Federica Bogo, Tony Tung, and Edmond Boyer.
\newblock Splatfields: Neural gaussian splats for sparse 3d and 4d reconstruction.
\newblock In \emph{European Conference on Computer Vision}, pages 313--332. Springer, 2024.

\bibitem[Mildenhall et~al.(2021)Mildenhall, Srinivasan, Tancik, Barron, Ramamoorthi, and Ng]{nerf}
Ben Mildenhall, Pratul~P Srinivasan, Matthew Tancik, Jonathan~T Barron, Ravi Ramamoorthi, and Ren Ng.
\newblock Nerf: Representing scenes as neural radiance fields for view synthesis.
\newblock \emph{Communications of the ACM}, 65\penalty0 (1):\penalty0 99--106, 2021.

\bibitem[Moon et~al.(2024)Moon, Shiratori, and Saito]{exavatar}
Gyeongsik Moon, Takaaki Shiratori, and Shunsuke Saito.
\newblock Expressive whole-body 3d gaussian avatar.
\newblock In \emph{European Conference on Computer Vision}, pages 19--35. Springer, 2024.

\bibitem[Moreau et~al.(2024)Moreau, Song, Dhamo, Shaw, Zhou, and P{\'e}rez-Pellitero]{humangaussiansplatting-moreau}
Arthur Moreau, Jifei Song, Helisa Dhamo, Richard Shaw, Yiren Zhou, and Eduardo P{\'e}rez-Pellitero.
\newblock Human gaussian splatting: Real-time rendering of animatable avatars.
\newblock In \emph{Proceedings of the IEEE/CVF conference on computer vision and pattern recognition}, pages 788--798, 2024.

\bibitem[Niemeyer et~al.(2024)Niemeyer, Manhardt, Rakotosaona, Oechsle, Duckworth, Gosula, Tateno, Bates, Kaeser, and Tombari]{radsplat}
Michael Niemeyer, Fabian Manhardt, Marie-Julie Rakotosaona, Michael Oechsle, Daniel Duckworth, Rama Gosula, Keisuke Tateno, John Bates, Dominik Kaeser, and Federico Tombari.
\newblock Radsplat: Radiance field-informed gaussian splatting for robust real-time rendering with 900+ fps.
\newblock \emph{arXiv preprint arXiv:2403.13806}, 2024.

\bibitem[Palandra et~al.(2024)Palandra, Sanchietti, Baieri, and Rodol{\`a}]{gsedit}
Francesco Palandra, Andrea Sanchietti, Daniele Baieri, and Emanuele Rodol{\`a}.
\newblock Gsedit: Efficient text-guided editing of 3d objects via gaussian splatting.
\newblock \emph{arXiv preprint arXiv:2403.05154}, 2024.

\bibitem[Papantonakis et~al.(2024)Papantonakis, Kopanas, Kerbl, Lanvin, and Drettakis]{reducedgs}
Panagiotis Papantonakis, Georgios Kopanas, Bernhard Kerbl, Alexandre Lanvin, and George Drettakis.
\newblock Reducing the memory footprint of 3d gaussian splatting.
\newblock \emph{Proceedings of the ACM on Computer Graphics and Interactive Techniques}, 7\penalty0 (1):\penalty0 1--17, 2024.

\bibitem[Qian et~al.(2025)Qian, Yan, Gao, Ge, Wei, Shangguan, and He]{c3dgs}
Jiating Qian, Yiming Yan, Fengjiao Gao, Baoyu Ge, Maosheng Wei, Boyi Shangguan, and Guangjun He.
\newblock C3dgs: Compressing 3d gaussian model for surface reconstruction of large-scale scenes based on multi-view uav images.
\newblock \emph{IEEE Journal of Selected Topics in Applied Earth Observations and Remote Sensing}, 2025.

\bibitem[Qian et~al.(2024)Qian, Wang, Mihajlovic, Geiger, and Tang]{3dgs-avatar}
Zhiyin Qian, Shaofei Wang, Marko Mihajlovic, Andreas Geiger, and Siyu Tang.
\newblock 3dgs-avatar: Animatable avatars via deformable 3d gaussian splatting.
\newblock In \emph{Proceedings of the IEEE/CVF conference on computer vision and pattern recognition}, pages 5020--5030, 2024.

\bibitem[Qu et~al.(2025)Qu, Chen, Li, Li, Zhang, Cao, and Ji]{dragyourgaussian}
Yansong Qu, Dian Chen, Xinyang Li, Xiaofan Li, Shengchuan Zhang, Liujuan Cao, and Rongrong Ji.
\newblock Drag your gaussian: Effective drag-based editing with score distillation for 3d gaussian splatting.
\newblock \emph{arXiv preprint arXiv:2501.18672}, 2025.

\bibitem[R{\"u}ckert et~al.(2022)R{\"u}ckert, Franke, and Stamminger]{adop}
Darius R{\"u}ckert, Linus Franke, and Marc Stamminger.
\newblock Adop: Approximate differentiable one-pixel point rendering.
\newblock \emph{ACM Transactions on Graphics (ToG)}, 41\penalty0 (4):\penalty0 1--14, 2022.

\bibitem[Shao et~al.(2024)Shao, Wang, Li, Wang, Lin, Zhang, Fan, and Wang]{splattingavatar}
Zhijing Shao, Zhaolong Wang, Zhuang Li, Duotun Wang, Xiangru Lin, Yu Zhang, Mingming Fan, and Zeyu Wang.
\newblock Splattingavatar: Realistic real-time human avatars with mesh-embedded gaussian splatting.
\newblock In \emph{Proceedings of the IEEE/CVF Conference on Computer Vision and Pattern Recognition}, pages 1606--1616, 2024.

\bibitem[Svitov et~al.(2024)Svitov, Morerio, Agapito, and Del~Bue]{haha}
David Svitov, Pietro Morerio, Lourdes Agapito, and Alessio Del~Bue.
\newblock Haha: Highly articulated gaussian human avatars with textured mesh prior.
\newblock In \emph{Proceedings of the Asian Conference on Computer Vision}, pages 4051--4068, 2024.

\bibitem[Tang et~al.(2023)Tang, Ren, Zhou, Liu, and Zeng]{dreamgaussian}
Jiaxiang Tang, Jiawei Ren, Hang Zhou, Ziwei Liu, and Gang Zeng.
\newblock Dreamgaussian: Generative gaussian splatting for efficient 3d content creation.
\newblock \emph{arXiv preprint arXiv:2309.16653}, 2023.

\bibitem[von L{\"u}tzow and Nie{\ss}ner(2025)]{linprim}
Nicolas von L{\"u}tzow and Matthias Nie{\ss}ner.
\newblock Linprim: Linear primitives for differentiable volumetric rendering.
\newblock \emph{arXiv preprint arXiv:2501.16312}, 2025.

\bibitem[Wang et~al.(2024)Wang, Zhu, He, Feng, Deng, Bian, and Chen]{rdgaussian}
Henan Wang, Hanxin Zhu, Tianyu He, Runsen Feng, Jiajun Deng, Jiang Bian, and Zhibo Chen.
\newblock End-to-end rate-distortion optimized 3d gaussian representation.
\newblock In \emph{European Conference on Computer Vision}, pages 76--92. Springer, 2024.

\bibitem[Wu et~al.(2024{\natexlab{a}})Wu, Yi, Fang, Xie, Zhang, Wei, Liu, Tian, and Wang]{4dsplat}
Guanjun Wu, Taoran Yi, Jiemin Fang, Lingxi Xie, Xiaopeng Zhang, Wei Wei, Wenyu Liu, Qi Tian, and Xinggang Wang.
\newblock 4d gaussian splatting for real-time dynamic scene rendering.
\newblock In \emph{Proceedings of the IEEE/CVF conference on computer vision and pattern recognition}, pages 20310--20320, 2024{\natexlab{a}}.

\bibitem[Wu et~al.(2024{\natexlab{b}})Wu, Bian, Li, Wang, Reid, Torr, and Prisacariu]{gaussctrl}
Jing Wu, Jia-Wang Bian, Xinghui Li, Guangrun Wang, Ian Reid, Philip Torr, and Victor~Adrian Prisacariu.
\newblock Gaussctrl: Multi-view consistent text-driven 3d gaussian splatting editing.
\newblock In \emph{European Conference on Computer Vision}, pages 55--71. Springer, 2024{\natexlab{b}}.

\bibitem[Xie et~al.(2024)Xie, Zong, Qiu, Li, Feng, Yang, and Jiang]{physgaussian}
Tianyi Xie, Zeshun Zong, Yuxing Qiu, Xuan Li, Yutao Feng, Yin Yang, and Chenfanfu Jiang.
\newblock Physgaussian: Physics-integrated 3d gaussians for generative dynamics.
\newblock In \emph{Proceedings of the IEEE/CVF Conference on Computer Vision and Pattern Recognition}, pages 4389--4398, 2024.

\bibitem[Xu et~al.(2024{\natexlab{a}})Xu, Ge, Qiu, Chen, Yan, Liu, Zhao, Zhao, Zhang, Liang, et~al.]{gaussianproperty}
Xinli Xu, Wenhang Ge, Dicong Qiu, ZhiFei Chen, Dongyu Yan, Zhuoyun Liu, Haoyu Zhao, Hanfeng Zhao, Shunsi Zhang, Junwei Liang, et~al.
\newblock Gaussianproperty: Integrating physical properties to 3d gaussians with lmms.
\newblock \emph{arXiv preprint arXiv:2412.11258}, 2024{\natexlab{a}}.

\bibitem[Xu et~al.(2024{\natexlab{b}})Xu, Jiang, Xiao, Feng, and Zhang]{dg-slam}
Yueming Xu, Haochen Jiang, Zhongyang Xiao, Jianfeng Feng, and Li Zhang.
\newblock Dg-slam: Robust dynamic gaussian splatting slam with hybrid pose optimization.
\newblock \emph{arXiv preprint arXiv:2411.08373}, 2024{\natexlab{b}}.

\bibitem[Xu et~al.(2025)Xu, Li, Zhang, Chen, Zhang, Leng, and Zhou]{fgs-slam}
Yansong Xu, Junlin Li, Wei Zhang, Siyu Chen, Shengyong Zhang, Yuquan Leng, and Weijia Zhou.
\newblock Fgs-slam: Fourier-based gaussian splatting for real-time slam with sparse and dense map fusion.
\newblock \emph{arXiv preprint arXiv:2503.01109}, 2025.

\bibitem[Yan et~al.(2024{\natexlab{a}})Yan, Qu, Xu, Zhao, Wang, Wang, and Li]{gs-slam}
Chi Yan, Delin Qu, Dan Xu, Bin Zhao, Zhigang Wang, Dong Wang, and Xuelong Li.
\newblock Gs-slam: Dense visual slam with 3d gaussian splatting.
\newblock In \emph{Proceedings of the IEEE/CVF Conference on Computer Vision and Pattern Recognition}, pages 19595--19604, 2024{\natexlab{a}}.

\bibitem[Yan et~al.(2024{\natexlab{b}})Yan, Li, Shao, Chen, Kai, Hwang, Zhao, and Remondino]{3dsceneeditor}
Ziyang Yan, Lei Li, Yihua Shao, Siyu Chen, Wuzong Kai, Jenq-Neng Hwang, Hao Zhao, and Fabio Remondino.
\newblock 3dsceneeditor: Controllable 3d scene editing with gaussian splatting.
\newblock \emph{arXiv preprint arXiv:2412.01583}, 2024{\natexlab{b}}.

\bibitem[Yang et~al.(2024)Yang, Gao, Zhou, Jiao, Zhang, and Jin]{deformable3d}
Ziyi Yang, Xinyu Gao, Wen Zhou, Shaohui Jiao, Yuqing Zhang, and Xiaogang Jin.
\newblock Deformable 3d gaussians for high-fidelity monocular dynamic scene reconstruction.
\newblock In \emph{Proceedings of the IEEE/CVF conference on computer vision and pattern recognition}, pages 20331--20341, 2024.

\bibitem[Ye et~al.(2024)Ye, Danelljan, Yu, and Ke]{gaussiangrouping}
Mingqiao Ye, Martin Danelljan, Fisher Yu, and Lei Ke.
\newblock Gaussian grouping: Segment and edit anything in 3d scenes.
\newblock In \emph{European Conference on Computer Vision}, pages 162--179. Springer, 2024.

\bibitem[Yuan et~al.(2024)Yuan, Li, Huang, De~Mello, Nagano, Kautz, and Iqbal]{gavatars}
Ye Yuan, Xueting Li, Yangyi Huang, Shalini De~Mello, Koki Nagano, Jan Kautz, and Umar Iqbal.
\newblock Gavatar: Animatable 3d gaussian avatars with implicit mesh learning.
\newblock In \emph{Proceedings of the IEEE/CVF Conference on Computer Vision and Pattern Recognition}, pages 896--905, 2024.

\bibitem[Zhang et~al.(2024{\natexlab{a}})Zhang, Wang, Wang, Li, Qin, and Wang]{gaussiansinthewild}
Dongbin Zhang, Chuming Wang, Weitao Wang, Peihao Li, Minghan Qin, and Haoqian Wang.
\newblock Gaussian in the wild: 3d gaussian splatting for unconstrained image collections.
\newblock In \emph{European Conference on Computer Vision}, pages 341--359. Springer, 2024{\natexlab{a}}.

\bibitem[Zhang et~al.(2024{\natexlab{b}})Zhang, Bi, Tan, Xiangli, Zhao, Sunkavalli, and Xu]{gs-lrm}
Kai Zhang, Sai Bi, Hao Tan, Yuanbo Xiangli, Nanxuan Zhao, Kalyan Sunkavalli, and Zexiang Xu.
\newblock Gs-lrm: Large reconstruction model for 3d gaussian splatting.
\newblock In \emph{European Conference on Computer Vision}, pages 1--19. Springer, 2024{\natexlab{b}}.

\bibitem[Zhang et~al.(2024{\natexlab{c}})Zhang, Ge, Xu, He, Wang, Qin, Lu, Geng, and Zhang]{gaussianimage}
Xinjie Zhang, Xingtong Ge, Tongda Xu, Dailan He, Yan Wang, Hongwei Qin, Guo Lu, Jing Geng, and Jun Zhang.
\newblock Gaussianimage: 1000 fps image representation and compression by 2d gaussian splatting.
\newblock In \emph{European Conference on Computer Vision}, 2024{\natexlab{c}}.

\bibitem[Zhang et~al.(2025)Zhang, Song, Lee, Yang, Peng, Chellappa, and Fan]{lp3dgs}
Zhaoliang Zhang, Tianchen Song, Yongjae Lee, Li Yang, Cheng Peng, Rama Chellappa, and Deliang Fan.
\newblock Lp-3dgs: Learning to prune 3d gaussian splatting.
\newblock \emph{Advances in Neural Information Processing Systems}, 37:\penalty0 122434--122457, 2025.

\bibitem[Zhong et~al.(2024)Zhong, Wang, Zhang, and Liao]{generativeobjectinsertion}
Hongliang Zhong, Can Wang, Jingbo Zhang, and Jing Liao.
\newblock Generative object insertion in gaussian splatting with a multi-view diffusion model.
\newblock \emph{arXiv preprint arXiv:2409.16938}, 2024.

\end{thebibliography}
}


\pagebreak
\appendix
\vfill 

\pagebreak

\section{Acceptance Probability Derivation}
\subsection{Proposal and Acceptance Probabilities}
\label{sec:appendix_acceptance}

To determine whether a split, merge or parameter update step proposed is accepted to be taken or not - we compute the acceptance probability as follows:
\begin{equation}
    A(S \rightarrow S') = \min\left(1, \frac{P(S') q(S | S')}{P(S) q
    (S' | S)}\right)
\end{equation}
For the parameter update transition that uses the SGLD update, $A=1$. For the merge and split transitions, the acceptance probability is given by:
\begin{equation}
A(S \rightarrow S_{\text{merge}}) = \min\left(1, \frac{P(S_{\text{merge}}) q_{\text{split}}(S_{\text{merge}}\rightarrow S)}{P(S) q_{\text{merge}}(S \rightarrow S_{\text{merge}})}\right)    
\end{equation}

\begin{equation}
A(S \rightarrow S_{\text{split}}) = \min\left(1, \frac{P(S_{\text{split}}) q_{\text{merge}}(S_{\text{split}}\rightarrow S)}{P(S) q_{\text{split}}(S \rightarrow S_{\text{split}})}\right)    \end{equation}

We now derive the acceptance distribution. 

\subsubsection{Parameter update choice} The proposal distribution for a parameter update, $q_{update}$, is identical to the SGLD parameter update. Thus we express the parameter update state transition using $q_{update}$ as follows:
\begin{equation}
\boldsymbol{p}_{t+1} \sim \mathcal{N}\left(\boldsymbol{p}_t + \frac{\epsilon_p}{2}\nabla_p \log (p (S)), \epsilon_p I\right)    
\end{equation}
\begin{equation}
o_{t+1} \sim \mathcal{N}\left(o_t + \frac{\epsilon_o}{2}\nabla_o \log (p (S)), \epsilon_o I\right)    
\end{equation}
\begin{equation}
\text{SH}_{t+1} \sim \mathcal{N}\left(\text{SH}_t + \frac{\epsilon_{\text{SH}}}{2}\nabla_{\text{SH}} \log (p(G, C)), \epsilon_{\text{SH}} I\right)    
\end{equation}
\begin{equation}   
\text{SR}_{t+1} \sim \mathcal{N}\left(\text{SR}_t + \frac{\epsilon_{\text{SR}}}{2}\nabla_{\text{SR}} \log (p (G, C)), \epsilon_{\text{SR}} I\right)
\end{equation}
Where $\epsilon_p, \epsilon_o, \epsilon_{\text{SH}}, \epsilon_{\text{SR}}$ are small hyperparameters. The acceptance probability $A=1$.

\subsubsection{Split Codebook Vectors}
\label{sec:split_vector_acceptance}
During the split transition, we split its codebook vector into a new entry. A 3DG mapped to codebook vector $c$ is remapped into a new codebook vector entry row $c''$. The original codebook vector becomes $c'$. 
\begin{equation}
c' = c \qquad c'' = c + u    
\end{equation}
\begin{equation}
q_{split}(u) = \mathcal{N}(0, \epsilon_{\text{split}}I)
\end{equation}
The acceptance probability of this state transition is given by:
\begin{equation}
A(S\rightarrow S_{\text{split}}) = \min\left(1, \frac{p(S_{\text{split}})q_{merge}(c-c')}{p(S)q_{split}(c-c')}\right)
\end{equation}
The ratio $\frac{p(S_{\text{split}})}{p(S)}\approx e^{-\lambda_{\text{SH}}}$ if we choose to split the SH codebook, and $e^{-\lambda_{\text{SR}}}$ if we choose to split the SR codebook. We thus have the acceptance probability given by:  
\begin{equation}
A(S \rightarrow S_{\text{split}}) = \min \left(1, e^{-\lambda_{\text{SH}}}/q_{sm}(u)\right)
\end{equation}

\subsubsection{Merge Codebook Vectors}
\label{sec:merge_vector_acceptance}
Two codebook vectors $c, c'$ can be merged into one codebook vector of value $c$. 
Two rows to be merged are selected with a transition distribution defined by:
\begin{equation}
q_{\text{merge}}(S\rightarrow S_{\text{merge}}) = \mathcal{N}(c-c'| \mu=0,\sigma=\epsilon_{\text{merge}} I)
\end{equation}
The acceptance probability of a merge transition is:
\begin{align}
\footnotesize	
A(S &\rightarrow S_{\text{merge}}) = \\\nonumber
\min&\left(1, \frac{p(S_{\text{merge}})q_{split}(c-c') }{p(S)q_{merge}(c-c')} \right)
\end{align}
$\frac{p(S)}{p(S_{\text{merge}})}\approx e^{\lambda_{\text{SH}}}$ if it leads to a reduction in the number of rows, or $1$ otherwise, as merging a small set of rows of codebook vectors does not affect the overall accuracy.
\begin{equation}
A(S \rightarrow S_{\text{merge}}) = \min\left(1,
e^{\lambda_{\text{SH}}}q_{sm}(c-c')\right)
\end{equation}
Where 
\begin{equation}
q_{sm}(u) = \exp\left( \frac{u^2}{2}\left(\frac{1}{\epsilon^2_{merge}} - \frac{1}{\epsilon^2_{split}}\right) \right)
\end{equation}

\vfill
\pagebreak

\onecolumn

\section{Reconstruction using an SfM Initialized Point Cloud}
\label{sec:appendix_sfmrecon}

\begin{table}[!htb]
\resizebox{\columnwidth}{!}{%
\begin{tabular}{
>{\columncolor[HTML]{F4F6F8}}l 
>{\columncolor[HTML]{FFFFFF}}c 
>{\columncolor[HTML]{FFFFFF}}c 
>{\columncolor[HTML]{FFFFFF}}c 
>{\columncolor[HTML]{FFFFFF}}c 
>{\columncolor[HTML]{FFFFFF}}c 
>{\columncolor[HTML]{FFFFFF}}c 
>{\columncolor[HTML]{FFFFFF}}c 
>{\columncolor[HTML]{FFFFFF}}c 
>{\columncolor[HTML]{FFFFFF}}c 
>{\columncolor[HTML]{FFFFFF}}c }
\cellcolor[HTML]{FFFFFF} & DeepBlending & \multicolumn{7}{c}{\cellcolor[HTML]{FFFFFF}MipNerf360} & \multicolumn{2}{c}{\cellcolor[HTML]{FFFFFF}Tanks and Temples} \\
\cellcolor[HTML]{FFFFFF} & playroom & bicycle & bonsai & counter & garden & kitchen & room & stump & train & truck \\
3DGS & 30.10 & \cellcolor{orange!25}25.18 & 31.98 & 29.13 & \cellcolor{yellow!25}27.38 & \cellcolor{orange!25}31.54 & 31.77 & \cellcolor{yellow!25}26.67 & 22.03 & 25.39 \\
EAGLES & \cellcolor{yellow!25}30.38 & 25.02 & 31.45 & 28.42 & 26.94 & 30.79 & 31.64 & \cellcolor{yellow!25}26.67 & 22.34 & 25.04 \\
Reduced-GS & 29.96 & \cellcolor{yellow!25}25.12 & \cellcolor{yellow!25}32.10 & 29.13 & 27.28 & 31.33 & 31.68 & 26.58 & 22.01 & 25.42 \\
Scaffold-GS & \cellcolor{orange!25}30.89 & 25.02 & \cellcolor{orange!25}32.50 & \cellcolor{yellow!25}29.43 & 27.30 & \cellcolor{yellow!25}31.42 & \cellcolor{yellow!25}32.13 & \cellcolor{orange!25}26.72 & \cellcolor{yellow!25}22.54 & \cellcolor{yellow!25}25.74 \\
SpeedySplat & 30.02 & 25.10 & 31.20 & 28.28 & 26.68 & 29.65 & 30.78 & 26.64 & 21.68 & 25.23 \\
Taming3DGS & 30.11 & 24.86 & \cellcolor{red!25}32.93 & \cellcolor{orange!25}29.63 & \cellcolor{red!25}27.59 & \cellcolor{red!25}32.16 & \cellcolor{red!25}32.40 & 26.17 & \cellcolor{orange!25}22.58 & \cellcolor{orange!25}26.00 \\
\X-2M & \cellcolor{red!25}33.73 & \cellcolor{red!25}27.02 & 31.59 & \cellcolor{red!25}30.60 & \cellcolor{orange!25}27.49 & 30.54 & \cellcolor{orange!25}32.34 & \cellcolor{red!25}30.83 & \cellcolor{red!25}24.35 & \cellcolor{red!25}26.93 \\
\end{tabular}%
}
\caption{PSNR on evaluation datasets}
\label{tab:sfm-psnr}
\end{table}

\begin{table}[!htb]
\resizebox{\columnwidth}{!}{%
\begin{tabular}{
>{\columncolor[HTML]{F4F6F8}}l 
>{\columncolor[HTML]{FFFFFF}}c 
>{\columncolor[HTML]{FFFFFF}}c 
>{\columncolor[HTML]{FFFFFF}}c 
>{\columncolor[HTML]{FFFFFF}}c 
>{\columncolor[HTML]{FFFFFF}}c 
>{\columncolor[HTML]{FFFFFF}}c 
>{\columncolor[HTML]{FFFFFF}}c 
>{\columncolor[HTML]{FFFFFF}}c 
>{\columncolor[HTML]{FFFFFF}}c 
>{\columncolor[HTML]{FFFFFF}}c }
\cellcolor[HTML]{FFFFFF} & DeepBlending & \multicolumn{7}{c}{\cellcolor[HTML]{FFFFFF}MipNerf360} & \multicolumn{2}{c}{\cellcolor[HTML]{FFFFFF}Tanks and Temples} \\
\cellcolor[HTML]{FFFFFF} & playroom & bicycle & bonsai & counter & garden & kitchen & room & stump & train & truck \\
3DGS & 0.909 & \cellcolor{yellow!25}0.748 & 0.946 & 0.916 & \cellcolor{red!25}0.858 & 0.916 & 0.916 & \cellcolor{yellow!25}0.768 & 0.821 & 0.885 \\
EAGLES & \cellcolor{orange!25}0.913 & \cellcolor{orange!25}0.750 & 0.942 & 0.907 & 0.840 & 0.928 & 0.927 & \cellcolor{orange!25}0.774 & 0.798 & 0.876 \\
Reduced-GS & 0.906 & 0.747 & \cellcolor{yellow!25}0.947 & 0.915 & \cellcolor{orange!25}0.856 & \cellcolor{orange!25}0.932 & 0.926 & \cellcolor{yellow!25}0.768 & 0.810 & 0.882 \\
Scaffold-GS & \cellcolor{orange!25}0.913 & 0.740 & \cellcolor{orange!25}0.948 & \cellcolor{yellow!25}0.917 & 0.850 & \cellcolor{yellow!25}0.929 & \cellcolor{yellow!25}0.931 & 0.766 & \cellcolor{yellow!25}0.829 & \cellcolor{yellow!25}0.887 \\
SpeedySplat & 0.907 & 0.747 & 0.926 & 0.877 & 0.815 & 0.891 & 0.904 & 0.764 & 0.773 & 0.868 \\
Taming3DGS & 0.910 & 0.706 & \cellcolor{red!25}0.950 & \cellcolor{orange!25}0.922 & \cellcolor{yellow!25}0.856 & \cellcolor{red!25}0.937 & \cellcolor{red!25}0.934 & 0.738 & \cellcolor{orange!25}0.830 & \cellcolor{orange!25}0.893 \\
\X-2M & \cellcolor{red!25}0.925 & \cellcolor{red!25}0.813 & 0.943 & \cellcolor{red!25}0.938 & 0.848 & 0.922 & \cellcolor{orange!25}0.933 & \cellcolor{red!25}0.896 & \cellcolor{red!25}0.861 & \cellcolor{red!25}0.904 \\
\end{tabular}%
}
\caption{SSIM on evaluation datasets}
\label{tab:sfm-ssim}
\end{table}

\begin{table}[!htb]
\resizebox{\columnwidth}{!}{%
\begin{tabular}{
>{\columncolor[HTML]{F4F6F8}}l 
>{\columncolor[HTML]{FFFFFF}}c 
>{\columncolor[HTML]{FFFFFF}}c 
>{\columncolor[HTML]{FFFFFF}}c 
>{\columncolor[HTML]{FFFFFF}}c 
>{\columncolor[HTML]{FFFFFF}}c 
>{\columncolor[HTML]{FFFFFF}}c 
>{\columncolor[HTML]{FFFFFF}}c 
>{\columncolor[HTML]{FFFFFF}}c 
>{\columncolor[HTML]{FFFFFF}}c 
>{\columncolor[HTML]{FFFFFF}}c }
\cellcolor[HTML]{FFFFFF} & DeepBlending & \multicolumn{7}{c}{\cellcolor[HTML]{FFFFFF}MipNerf360} & \multicolumn{2}{c}{\cellcolor[HTML]{FFFFFF}Tanks and Temples} \\
\cellcolor[HTML]{FFFFFF} & playroom & bicycle & bonsai & counter & garden & kitchen & room & stump & train & truck \\
3DGS & \cellcolor{red!25}0.241 & \cellcolor{orange!25}0.242 & \cellcolor{yellow!25}0.180 & \cellcolor{yellow!25}0.182 & \cellcolor{red!25}0.122 & \cellcolor{orange!25}0.116 & 0.196 & \cellcolor{orange!25}0.242 & 0.197 & 0.141 \\
EAGLES & 0.251 & 0.244 & 0.191 & 0.199 & 0.154 & 0.127 & 0.200 & \cellcolor{yellow!25}0.243 & 0.237 & 0.164 \\
Reduced-GS & \cellcolor{orange!25}0.243 & \cellcolor{yellow!25}0.244 & \cellcolor{yellow!25}0.180 & 0.183 & \cellcolor{orange!25}0.123 & \cellcolor{yellow!25}0.117 & 0.197 & \cellcolor{yellow!25}0.243 & 0.206 & 0.147 \\
Scaffold-GS & \cellcolor{yellow!25}0.244 & 0.260 & \cellcolor{orange!25}0.179 & 0.185 & 0.133 & 0.122 & \cellcolor{orange!25}0.187 & 0.261 & \cellcolor{orange!25}0.190 & \cellcolor{yellow!25}0.136 \\
SpeedySplat & 0.269 & \cellcolor{yellow!25}0.244 & 0.227 & 0.258 & 0.213 & 0.197 & 0.257 & 0.289 & 0.289 & 0.190 \\
Taming3DGS & 0.251 & 0.314 & \cellcolor{red!25}0.172 & \cellcolor{orange!25}0.171 & \cellcolor{yellow!25}0.129 & \cellcolor{red!25}0.110 & \cellcolor{red!25}0.181 & 0.309 & \cellcolor{yellow!25}0.191 & \cellcolor{orange!25}0.125 \\
\X-2M & 0.252 & \cellcolor{red!25}0.230 & 0.186 & \cellcolor{red!25}0.150 & 0.160 & 0.148 & \cellcolor{yellow!25}0.187 & \cellcolor{red!25}0.163 & \cellcolor{red!25}0.187 & \cellcolor{red!25}0.124 \\
\end{tabular}%
}
\caption{LPIPS on evaluation datasets}
\label{tab:sfm-lpips}
\end{table}

\begin{table}[!htb]
\resizebox{\columnwidth}{!}{%
\begin{tabular}{lcccccccccccccccccc}
\rowcolor[HTML]{FFFFFF} 
 & \multicolumn{8}{c}{\cellcolor[HTML]{FFFFFF}Blender} & DeepBlending & \multicolumn{7}{c}{\cellcolor[HTML]{FFFFFF}MipNerf360} & \multicolumn{2}{c}{\cellcolor[HTML]{FFFFFF}Tanks and Temples} \\
\rowcolor[HTML]{FFFFFF} 
\textit{} & chair & drums & ficus & hotdog & lego & materials & mic & ship & playroom & bicycle & bonsai & counter & garden & kitchen & room & stump & train & truck \\
\cellcolor[HTML]{F4F6F8}\X-2M & \cellcolor[HTML]{FFFFFF}158.30 & \cellcolor[HTML]{FFFFFF}127.17 & \cellcolor[HTML]{FFFFFF}236.28 & \cellcolor[HTML]{FFFFFF}- & \cellcolor[HTML]{FFFFFF}103.08 & \cellcolor[HTML]{FFFFFF}149.54 & \cellcolor[HTML]{FFFFFF}257.87 & \cellcolor[HTML]{FFFFFF}115.33 & \cellcolor{yellow!25}134.57 & \cellcolor{red!25}130.04 & \cellcolor{orange!25}101.38 & \cellcolor{yellow!25}99.03 & \cellcolor{orange!25}139.62 & \cellcolor{yellow!25}135.20 & \cellcolor{yellow!25}128.97 & \cellcolor{red!25}131 & \cellcolor{yellow!25}128.07 & \cellcolor{orange!25}96.07 \\
\rowcolor[HTML]{FFFFFF} 
\cellcolor[HTML]{F4F6F8}3DGS & 190.27 & 149.03 & 67.31 & 70.85 & 140.61 & 58.35 & 83.00 & 115.99 & 817.16 & 2,089.36 & 482.08 & 467.02 & 1,858.58 & 690.12 & 553.89 & 1,772.67 & 480.17 & 901.40 \\
\rowcolor{red!25}
\cellcolor[HTML]{F4F6F8}EAGLES & 10.78 & 8.44 & 4.80 & 6.38 & 12.35 & 4.80 & 5.17 & 7.36 & 63.14 & \cellcolor{orange!25}159.47 & 50.91 & 44.77 & 116.03 & 82.74 & 52.33 & \cellcolor{orange!25}156.55 & 33.40 & 57.96 \\
\rowcolor[HTML]{FFFFFF} 
\cellcolor[HTML]{F4F6F8}Reduced-GS & 121.14 & 96.92 & 65.94 & 46.87 & 85.20 & 40.21 & 49.72 & 68.60 & 572.14 & 1,432.41 & 321.76 & 295.36 & 1,406.55 & 443.20 & 371.62 & 1,099.43 & 269.20 & 636.50 \\
\rowcolor{orange!25}
\cellcolor[HTML]{F4F6F8}Scaffold-GS & 31.66 & 31.66 & 31.66 & 31.66 & 31.66 & 31.66 & 31.66 & 31.96 & 95.11 & \cellcolor{yellow!25}304.83 & \cellcolor{yellow!25}135.45 & 91.52 & \cellcolor{yellow!25}246.28 & 107.76 & 89.25 & \cellcolor{yellow!25}252.44 & 114.89 & \cellcolor{yellow!25}192.21 \\
\rowcolor[HTML]{FFFFFF} 
\cellcolor[HTML]{F4F6F8}SpeedySplat & \cellcolor{yellow!25}72.30 & \cellcolor{yellow!25}57.09 & \cellcolor{yellow!25}43.51 & \cellcolor{yellow!25}32.94 & \cellcolor{yellow!25}61.88 & \cellcolor{yellow!25}32.71 & \cellcolor{yellow!25}32.45 & \cellcolor{yellow!25}50.27 & 393.28 & 895.20 & 314.86 & 264.14 & 1,043.39 & 406.80 & 286.05 & 803.08 & 188.92 & 494.39 \\
\rowcolor[HTML]{FFFFFF} 
\cellcolor[HTML]{F4F6F8}Taming3DGS & - & - & - & - & - & - & - & - & 325.63 & 477.60 & - & 631.04 & 1,221.08 & 797.98 & 756.10 & 282.02 & 743.60 & 1,189.74
\end{tabular}%
}
\caption{Peak memory during training}
\label{tab:peak_mem}
\end{table}

\pagebreak
\section{Random Parameter Initialization}
\label{sec:appendix_randompc}

\begin{table}[!htb]
\resizebox{\columnwidth}{!}{%
\begin{tabular}{lccccccccccccccccc}
\rowcolor[HTML]{FFFFFF} 
 & \multicolumn{2}{c}{\cellcolor[HTML]{FFFFFF}Tanks and Temples} & \multicolumn{7}{c}{\cellcolor[HTML]{FFFFFF}MipNerf360} & DeepBlending & \multicolumn{7}{c}{\cellcolor[HTML]{FFFFFF}Blender} \\
\rowcolor[HTML]{FFFFFF} 
 & train & truck & bicycle & bonsai & counter & garden & kitchen & room & stump & playroom & chair & drums & ficus & hotdog & lego & materials & mic \\
\rowcolor[HTML]{E06666} 
\rowcolor[HTML]{FFFFFF} 
\cellcolor[HTML]{F4F6F8}3DGS & 19.49 & 18.41 & 17.69 & 16.46 & 23.46 & 21.71 & 24.61 & \cellcolor{orange!25}27.69 & \cellcolor{yellow!25}20.74 & \cellcolor{yellow!25}14.41 & 31.93 & 24.91 & 29.05 & 36.48 & 32.37 & 29.69 & 34.58 \\
\rowcolor[HTML]{FFFFFF} 
\cellcolor[HTML]{F4F6F8}EAGLES & 18.91 & \cellcolor{yellow!25}18.43 & 19.30 & 17.79 & 23.25 & \cellcolor{orange!25}24.82 & 25.52 & \cellcolor{yellow!25}25.27 & 20.08 & \cellcolor{orange!25}20.82 & 34.42 & 25.68 & 33.67 & \cellcolor{yellow!25}37.05 & 34.92 & 28.95 & 35.12 \\
\cellcolor[HTML]{F4F6F8}Reduced-GS & \cellcolor[HTML]{FFFFFF}18.74 & \cellcolor[HTML]{FFFFFF}18.02 & \cellcolor[HTML]{FFFFFF} & \cellcolor[HTML]{FFFFFF}15.59 & \cellcolor{orange!25}23.85 & \cellcolor{yellow!25}24.15 & \cellcolor{yellow!25}25.82 & \cellcolor[HTML]{FFFFFF}25.11 & \cellcolor[HTML]{FFFFFF}19.67 & \cellcolor[HTML]{FFFFFF}12.95 & \cellcolor{orange!25}35.59 & \cellcolor{orange!25}26.28 & \cellcolor{orange!25}35.48 & \cellcolor{red!25}38.07 & \cellcolor{orange!25}36.06 & \cellcolor{yellow!25}30.50 & \cellcolor{orange!25}36.71 \\
\rowcolor[HTML]{FFFFFF} 
\cellcolor[HTML]{F4F6F8}Scaffold-GS & \cellcolor{yellow!25}19.69 & 18.32 & \cellcolor{orange!25}20.58 & 17.77 & 21.06 & 18.45 & 23.08 & 23.35 & 18.73 & 13.99 & \cellcolor{yellow!25}34.85 & \cellcolor{yellow!25}26.17 & 35.04 & \cellcolor{orange!25}37.81 & \cellcolor{yellow!25}35.42 & \cellcolor{orange!25}30.60 & \cellcolor{yellow!25}36.69 \\
\rowcolor[HTML]{FFFFFF} 
\cellcolor[HTML]{F4F6F8}SpeedySplat & 19.24 & 18.15 & 15.90 & \cellcolor{yellow!25}19.54 & \cellcolor{yellow!25}23.84 & 0.00 & 0.00 & 24.40 & 0.00 & 13.39 & 33.95 & 25.96 & \cellcolor{yellow!25}35.18 & 36.13 & 32.12 & 29.33 & 35.86 \\
\rowcolor[HTML]{FFFFFF} 
\cellcolor[HTML]{F4F6F8}Taming3DGS & \cellcolor{orange!25}19.78 & \cellcolor{orange!25}18.46 & \cellcolor{yellow!25}19.64 & \cellcolor{orange!25}20.08 & - & - & \cellcolor{orange!25}26.73 & - & \cellcolor{orange!25}20.86 & - & - & - & - & - & - & - & - \\
\rowcolor{red!25}
\cellcolor[HTML]{F4F6F8}\X-2M & 24.35 & 26.93 & 27.02 & 31.59 & 30.60 & 27.49 & 30.54 & 32.34 & 30.83 & 33.73 & 38.48 & 29.05 & 38.79 & \cellcolor[HTML]{FFFFFF}36.48 & 39.34 & 36.20 & 41.56 \\
\end{tabular}%
}
\caption{PSNR measured on evaluation dataset}
\label{tab:random_psnr}
\end{table}

\begin{table}[!htb]
\resizebox{\columnwidth}{!}{%
\begin{tabular}{lccccccccccccccccc}
\rowcolor[HTML]{FFFFFF} 
 & \multicolumn{2}{c}{\cellcolor[HTML]{FFFFFF}Tanks and Temples} & \multicolumn{7}{c}{\cellcolor[HTML]{FFFFFF}MipNerf360} & DeepBlending & \multicolumn{7}{c}{\cellcolor[HTML]{FFFFFF}Blender} \\
\rowcolor[HTML]{FFFFFF} 
 & train & truck & bicycle & bonsai & counter & garden & kitchen & room & stump & playroom & chair & drums & ficus & hotdog & lego & materials & mic \\
\rowcolor[HTML]{E06666} 
\rowcolor[HTML]{FFFFFF} 
\cellcolor[HTML]{F4F6F8}3DGS & \cellcolor{yellow!25}0.737 & \cellcolor{orange!25}0.722 & 0.487 & 0.658 & 0.827 & 0.710 & 0.878 & \cellcolor{orange!25}0.870 & \cellcolor{orange!25}0.606 & \cellcolor{yellow!25}0.682 & 0.983 & 0.941 & 0.953 & \cellcolor{orange!25}0.984 & 0.975 & 0.950 & 0.987 \\
\rowcolor[HTML]{FFFFFF} 
\cellcolor[HTML]{F4F6F8}EAGLES & 0.726 & 0.719 & \cellcolor{orange!25}0.562 & 0.690 & \cellcolor{yellow!25}0.835 & \cellcolor{orange!25}0.784 & 0.890 & \cellcolor{yellow!25}0.846 & \cellcolor{yellow!25}0.600 & \cellcolor{orange!25}0.808 & 0.984 & \cellcolor{yellow!25}0.950 & 0.982 & 0.983 & \cellcolor{yellow!25}0.980 & 0.949 & 0.989 \\
\cellcolor[HTML]{F4F6F8}Reduced-GS & \cellcolor[HTML]{FFFFFF}0.717 & \cellcolor[HTML]{FFFFFF}0.712 & \cellcolor[HTML]{FFFFFF}- & \cellcolor[HTML]{FFFFFF}0.620 & \cellcolor{orange!25}0.840 & \cellcolor{yellow!25}0.762 & \cellcolor{yellow!25}0.891 & \cellcolor[HTML]{FFFFFF}0.838 & \cellcolor[HTML]{FFFFFF}0.567 & \cellcolor[HTML]{FFFFFF}0.646 & \cellcolor{orange!25}0.988 & \cellcolor{orange!25}0.955 & \cellcolor{orange!25}0.987 & \cellcolor{red!25}0.985 & \cellcolor{orange!25}0.983 & \cellcolor{orange!25}0.960 & \cellcolor{orange!25}0.992 \\
\rowcolor[HTML]{FFFFFF} 
\cellcolor[HTML]{F4F6F8}Scaffold-GS & 0.727 & 0.696 & 0.467 & 0.697 & 0.780 & 0.455 & 0.830 & 0.803 & 0.396 & 0.669 & \cellcolor{yellow!25}0.985 & 0.948 & 0.985 & 0.984 & \cellcolor{yellow!25}0.980 & \cellcolor{yellow!25}0.960 & \cellcolor{yellow!25}0.992 \\
\rowcolor[HTML]{FFFFFF} 
\cellcolor[HTML]{F4F6F8}SpeedySplat & 0.706 & 0.701 & 0.422 & \cellcolor{yellow!25}0.713 & 0.816 & - & - & 0.817 & - & 0.657 & 0.979 & 0.949 & \cellcolor{yellow!25}0.985 & 0.975 & 0.958 & 0.947 & 0.990 \\
\rowcolor[HTML]{FFFFFF} 
\cellcolor[HTML]{F4F6F8}Taming3DGS & \cellcolor{orange!25}0.743 & \cellcolor{orange!25}0.722 & \cellcolor{yellow!25}0.525 & \cellcolor{orange!25}0.735 & - & - & \cellcolor{orange!25}0.895 & 0.000 & 0.588 & - & - & - & - & - & - & - & - \\
\rowcolor{red!25}
\cellcolor[HTML]{F4F6F8}\X-2M & 0.861 & 0.904 & 0.813 & 0.943 & 0.938 & 0.848 & 0.922 & 0.933 & 0.896 & 0.925 & 0.993 & 0.977 & 0.994 & \cellcolor{orange!25}0.984 & 0.992 & 0.989 & 0.997 \\
\end{tabular}%
}
\caption{SSIM measured on evaluation dataset}
\label{tab:random_ssim}
\end{table}

\begin{table}[!htb]
\resizebox{\columnwidth}{!}{%
\begin{tabular}{lccccccccccccccccc}
\rowcolor[HTML]{FFFFFF} 
 & \multicolumn{2}{c}{\cellcolor[HTML]{FFFFFF}Tanks and Temples} & \multicolumn{7}{c}{\cellcolor[HTML]{FFFFFF}MipNerf360} & DeepBlending & \multicolumn{7}{c}{\cellcolor[HTML]{FFFFFF}Blender} \\
\rowcolor[HTML]{FFFFFF} 
 & train & truck & bicycle & bonsai & counter & garden & kitchen & room & stump & playroom & chair & drums & ficus & hotdog & lego & materials & mic \\
\rowcolor[HTML]{FFFFFF} 
\cellcolor[HTML]{F4F6F8}3DGS & \cellcolor{yellow!25}0.286 & \cellcolor{orange!25}0.291 & 0.481 & 0.469 & 0.289 & 0.251 & 0.182 & \cellcolor{orange!25}0.279 & \cellcolor{orange!25}0.398 & 0.549 & 0.023 & 0.059 & 0.043 & 0.030 & 0.031 & 0.064 & 0.026 \\
\rowcolor[HTML]{FFFFFF} 
\cellcolor[HTML]{F4F6F8}EAGLES & 0.306 & 0.305 & \cellcolor{orange!25}0.422 & 0.443 & \cellcolor{orange!25}0.273 & \cellcolor{orange!25}0.203 & 0.174 & \cellcolor{yellow!25}0.312 & \cellcolor{yellow!25}0.406 & \cellcolor{yellow!25}0.387 & 0.015 & \cellcolor{yellow!25}0.045 & 0.018 & \cellcolor{yellow!25}0.025 & 0.020 & 0.053 & 0.011 \\
\cellcolor[HTML]{F4F6F8}Reduced-GS & \cellcolor[HTML]{FFFFFF}0.307 & \cellcolor[HTML]{FFFFFF}0.308 & \cellcolor[HTML]{FFFFFF}- & \cellcolor[HTML]{FFFFFF}0.498 & \cellcolor{yellow!25}0.277 & \cellcolor{yellow!25}0.219 & \cellcolor{yellow!25}0.168 & \cellcolor[HTML]{FFFFFF}0.320 & \cellcolor[HTML]{FFFFFF}0.441 & \cellcolor{orange!25}0.308 & \cellcolor{orange!25}0.010 & \cellcolor{orange!25}0.036 & \cellcolor{orange!25}0.012 & \cellcolor{red!25}0.020 & \cellcolor{orange!25}0.016 & \cellcolor{orange!25}0.037 & \cellcolor{orange!25}0.006 \\
\rowcolor[HTML]{FFFFFF} 
\cellcolor[HTML]{F4F6F8}Scaffold-GS & 0.291 & 0.317 & 0.505 & 0.456 & 0.353 & 0.508 & 0.262 & 0.368 & 0.571 & 0.573 & \cellcolor{yellow!25}0.014 & 0.047 & \cellcolor{yellow!25}0.014 & \cellcolor{orange!25}0.023 & \cellcolor{yellow!25}0.019 & \cellcolor{yellow!25}0.041 & \cellcolor{yellow!25}0.008 \\
\rowcolor[HTML]{FFFFFF} 
\cellcolor[HTML]{F4F6F8}SpeedySplat & 0.349 & 0.359 & 0.567 & \cellcolor{yellow!25}0.427 & 0.327 & - & - & 0.352 & 0.000 & 0.581 & 0.023 & 0.048 & 0.014 & 0.044 & 0.060 & 0.061 & 0.011 \\
\rowcolor[HTML]{FFFFFF} 
\cellcolor[HTML]{F4F6F8}Taming3DGS & \cellcolor{orange!25}0.282 & \cellcolor{yellow!25}0.292 & \cellcolor{yellow!25}0.454 & \cellcolor{orange!25}0.404 & - & - & \cellcolor{orange!25}0.161 & - & 0.406 & - & - & - & - & - & - & - & - \\
\rowcolor{red!25} 
\cellcolor[HTML]{F4F6F8}\X-2M & 0.187 & 0.124 & 0.230 & 0.186 & 0.150 & 0.160 & 0.148 & 0.187 & 0.163 & 0.252 & 0.007 & 0.026 & 0.007 & \cellcolor[HTML]{FFFFFF}0.030 & 0.009 & 0.019 & 0.003 \\
\end{tabular}%
}
\caption{LPIPS measured on evaluation dataset}
\label{tab:random_lpips}
\end{table}

\pagebreak

\section{Training and Rendering Speeds}
\label{sec:appendix_train_renderingspeeds}

\begin{table}[!htb]
\resizebox{\columnwidth}{!}{%
\begin{tabular}{
>{\columncolor[HTML]{F4F6F8}}l 
>{\columncolor[HTML]{FFFFFF}}c 
>{\columncolor[HTML]{FFFFFF}}c 
>{\columncolor[HTML]{FFFFFF}}c 
>{\columncolor[HTML]{FFFFFF}}c 
>{\columncolor[HTML]{FFFFFF}}c 
>{\columncolor[HTML]{FFFFFF}}c 
>{\columncolor[HTML]{FFFFFF}}c 
>{\columncolor[HTML]{FFFFFF}}c 
>{\columncolor[HTML]{FFFFFF}}c 
>{\columncolor[HTML]{FFFFFF}}c }
\cellcolor[HTML]{FFFFFF} & \multicolumn{2}{c}{\cellcolor[HTML]{FFFFFF}T and T} & \multicolumn{7}{c}{\cellcolor[HTML]{FFFFFF}MipNerf360} & DeepBlending \\
\cellcolor[HTML]{FFFFFF}\textit{} & train & truck & bicycle & bonsai & counter & garden & kitchen & room & stump & playroom \\
Taming3DGS & 263.40 & 240.66 & 224.62 & 211.51 & 170.02 & 178.84 & 163.10 & 174.79 & 374.40 & 468.09 \\
Ours 5M & 129.00 & 96.00 & 89.00 & 58.00 & 46.00 & 118.00 & 101.00 & 107.00 & 88.00 & 141.00 \\
Ours & 206.00 & 209.00 & 159.00 & 217.00 & 219.00 & 249.19 & 210.00 & 242.00 & 220.00 & 319.00 \\
MCMC 5M & 75.50 & 85.90 & 72.00 & 68.50 & 56.10 & 91.70 & 59.49 & 71.24 & 41.00 & 114.00 \\
MCMC 2M & 94.00 & 173.32 & 77.00 & 75.69 & 62.78 & 185.00 & 70.16 & 143.58 & 142.79 & 251.8
\end{tabular}%
}
\caption{Frames per second (FPS) measured on differnt datasets}
\label{tab:fps}
\end{table}

\begin{table}[!htb]
\resizebox{\columnwidth}{!}{%
\begin{tabular}{
>{\columncolor[HTML]{F4F6F8}}l 
>{\columncolor[HTML]{FFFFFF}}c 
>{\columncolor[HTML]{FFFFFF}}c 
>{\columncolor[HTML]{FFFFFF}}c 
>{\columncolor[HTML]{FFFFFF}}c 
>{\columncolor[HTML]{FFFFFF}}c 
>{\columncolor[HTML]{FFFFFF}}c 
>{\columncolor[HTML]{FFFFFF}}c 
>{\columncolor[HTML]{FFFFFF}}c 
>{\columncolor[HTML]{FFFFFF}}c 
>{\columncolor[HTML]{FFFFFF}}c }
\cellcolor[HTML]{FFFFFF} & DeepBlending & \multicolumn{7}{c}{\cellcolor[HTML]{FFFFFF}MipNerf360} & \multicolumn{2}{c}{\cellcolor[HTML]{FFFFFF}T and T} \\
\cellcolor[HTML]{FFFFFF}\textit{} & playroom & bicycle & bonsai & counter & garden & kitchen & room & stump & train & truck \\
MCMC 2M & 31.66 & 24.00 & 24.00 & 20.00 & 27.00 & 22.87 & 23.70 & 24.00 & 29.19 & 31.49 \\
MCMC 5M & 13.10 & 11.99 & 11.99 & 10.09 & 13.08 & 10.81 & 11.28 & 12.76 & 12.25 & 13.65 \\
Ours & 46.33 & 31.63 & 27.07 & 32.82 & 36.33 & 29.48 & 34.80 & 32.00 & 38.50 & 43.00 \\
Ours 5M & 21.65 & 15.09 & 13.55 & 11.59 & 17.80 & 13.14 & 17.45 & 17.33 & 19.87 & 18.48 \\
Taming3DGS & 93.12 & 71.35 & 50.09 & 48.59 & 40.02 & 40.70 & 49.09 & 102.78 & 67.08 & 47.11
\end{tabular}%
}
\caption{Training Iterations per second}
\label{tab:train_it_s}
\end{table}

\subsection{Comparison of Training Speeds with Taming-GS at Different Gaussian Counts}
Table~\ref{tab:taminggs_contrags} compares training speeds with TamingGS.

\begin{table}[!htb]
    \vspace{-0.3cm}
    \centering
    \begin{tabular}{|c|c|c|c|}

    \hline
         Model-numGS & PSNR & Training time & Memory\\
    \hline
         ContraGS-2M & 30.06 & 14 mins & 130 MB \\
         ContraGS-530K & 29.31 & 6.5 mins & 59 MB \\
         TamingGS & 29.39 &  8 mins & 477 MB \\
    \hline
    \end{tabular}
    \vspace{-0.2cm}
    \caption{Performance comparison of TamingGS and ContraGS for the same number of Gaussians}
    \label{tab:taminggs_contrags}
    \vspace{-0.3cm}
\end{table}

\pagebreak

\section{Qualitative Results}
\label{sec:qualtative}


\begin{figure}[!htb]
    \centering
    \footnotesize
    \setlength\tabcolsep{0pt}
    \begin{tabularx}{0.92\linewidth}{p{1em} *{2}{>{\centering\arraybackslash}X}} 
        & Ground Truth & Ours \\

        {\rotatebox{90}{\,\,\, Bicycle}}&
        \includegraphics[width=0.98\linewidth]{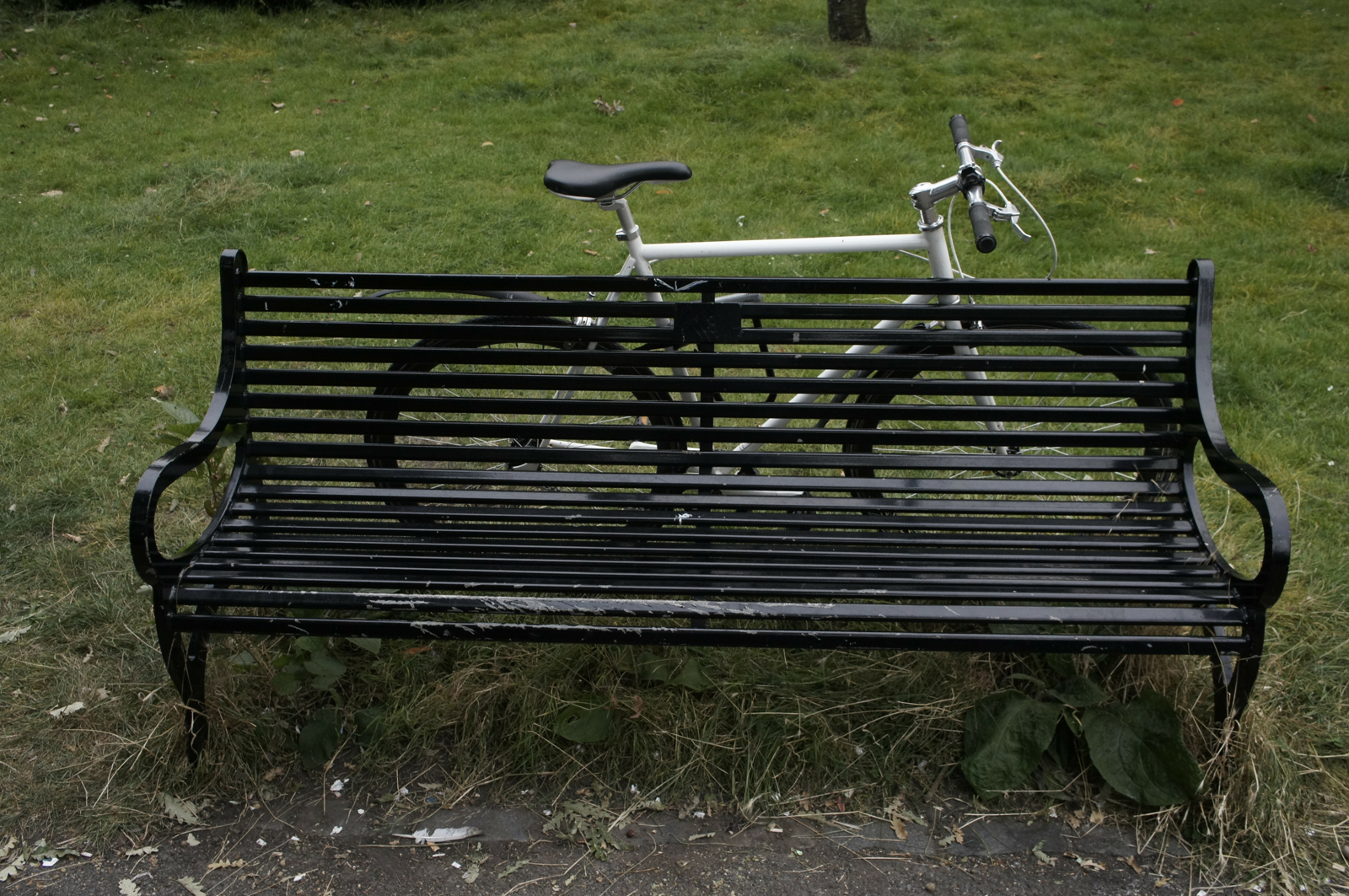} &
        \includegraphics[width=0.98\linewidth]{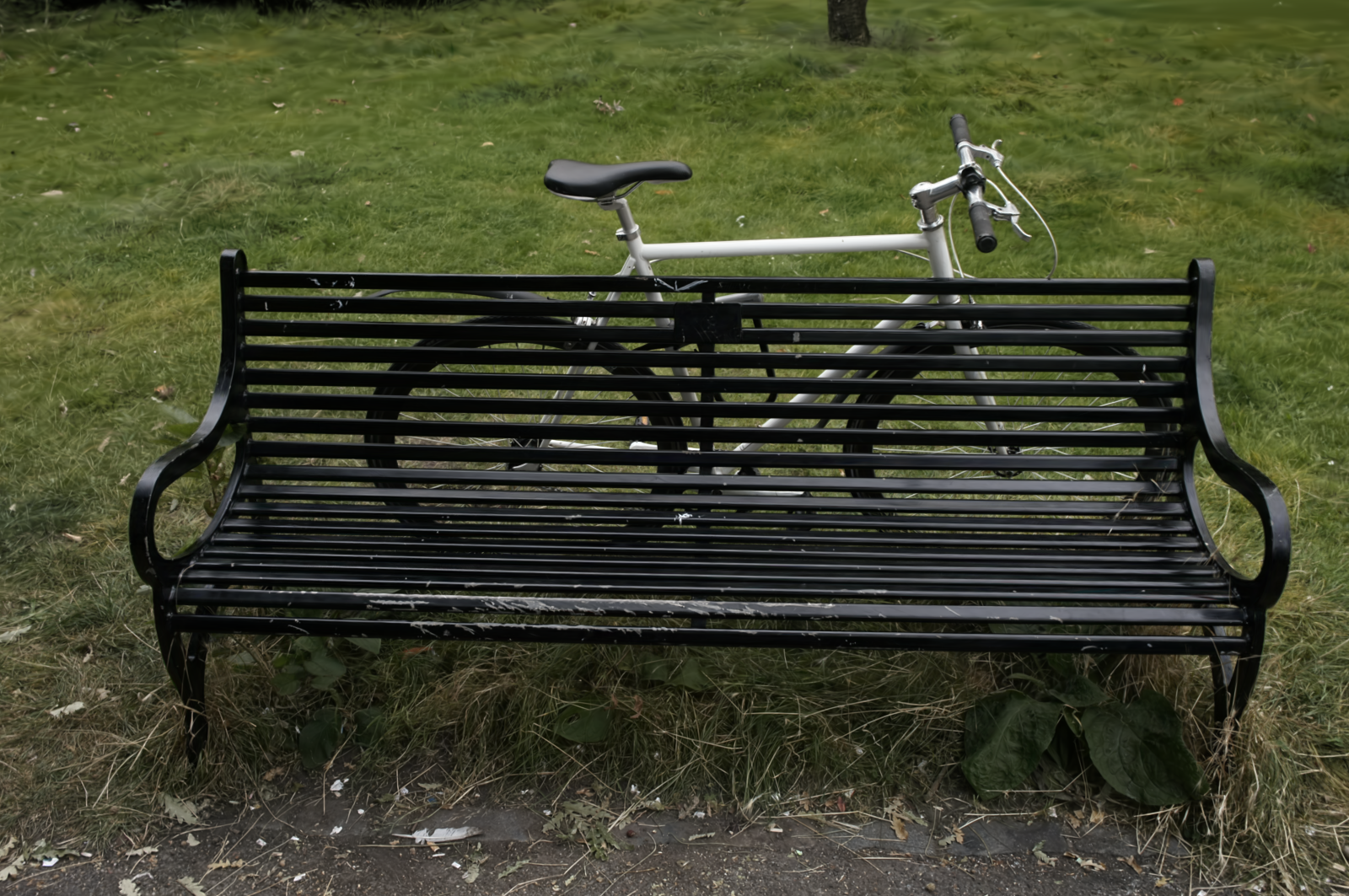} \\

        {\rotatebox{90}{\,\,\, Stump}}&
        \includegraphics[width=0.98\linewidth]{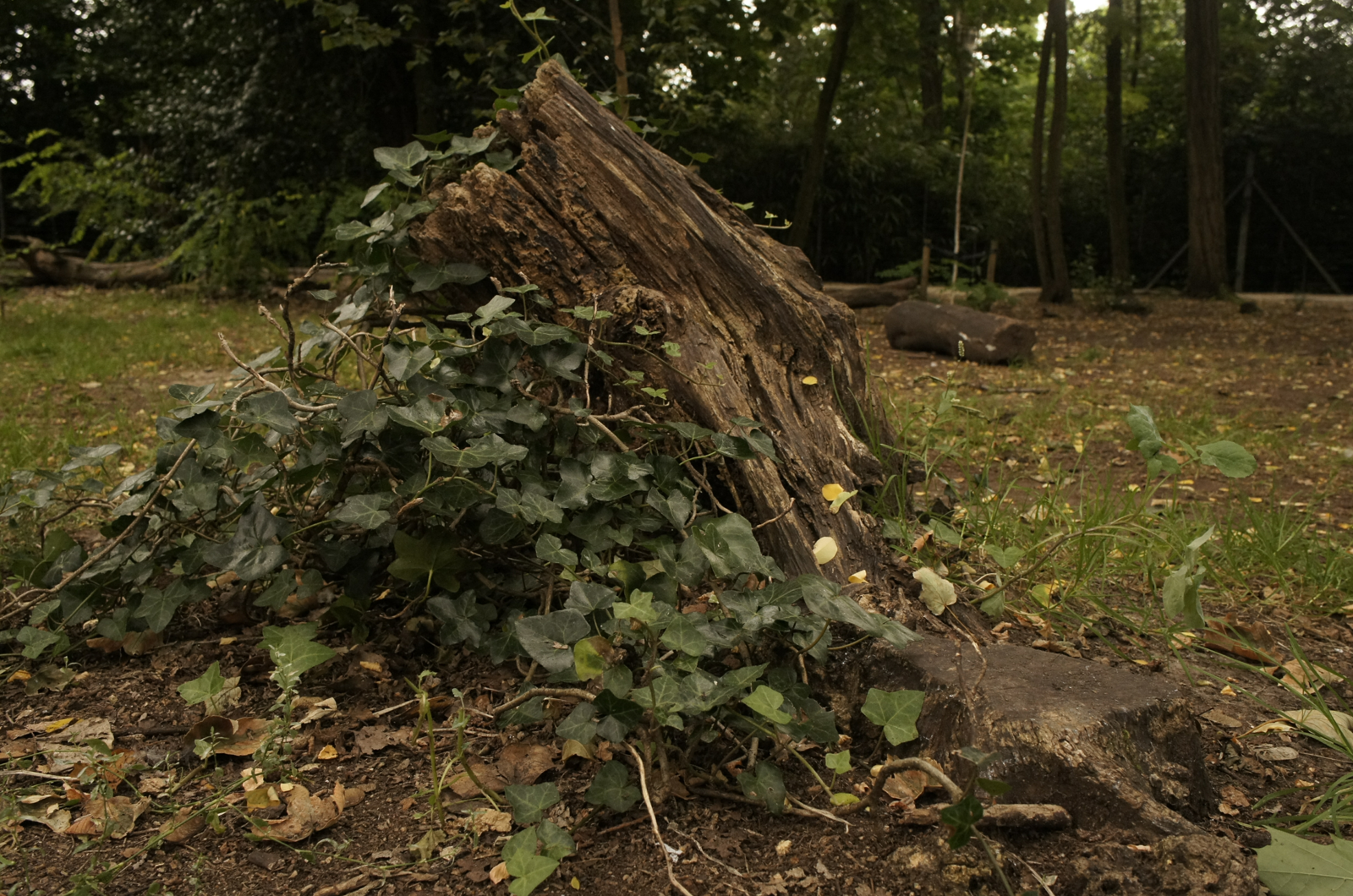} &
        \includegraphics[width=0.98\linewidth]{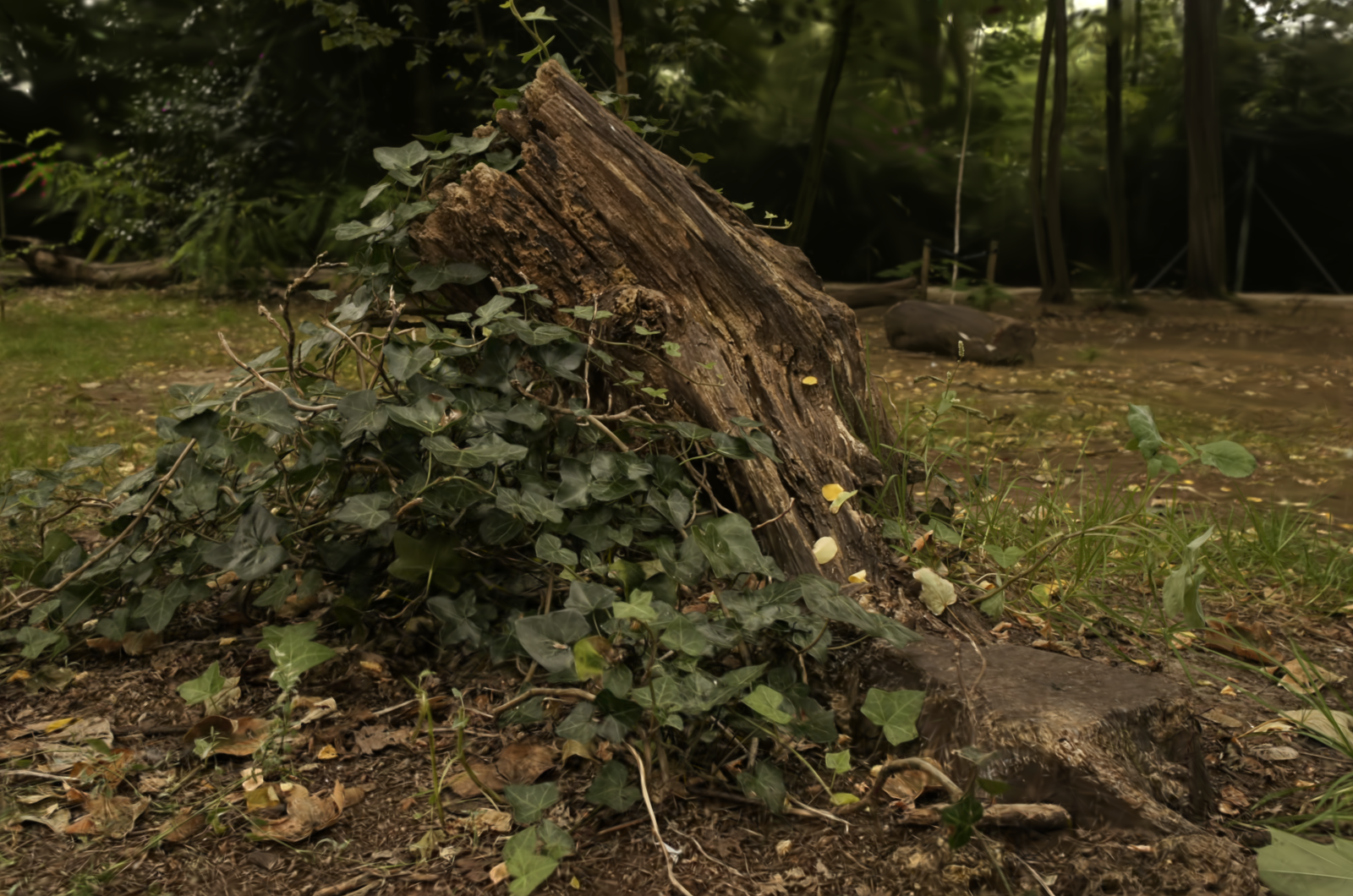} \\

        {\rotatebox{90}{\,\,\, Train}}&
        \includegraphics[width=0.98\linewidth]{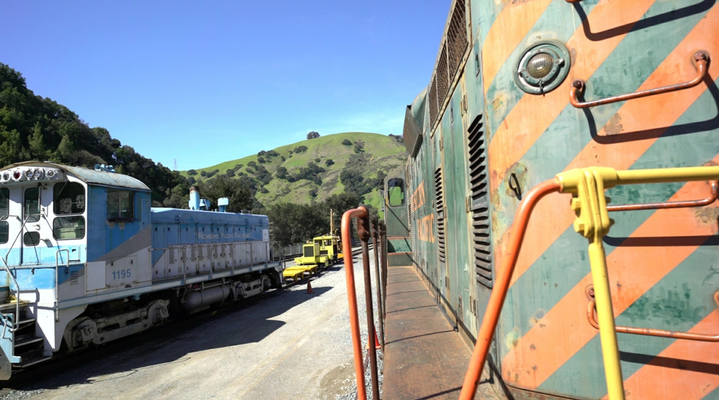} &
        \includegraphics[width=0.98\linewidth]{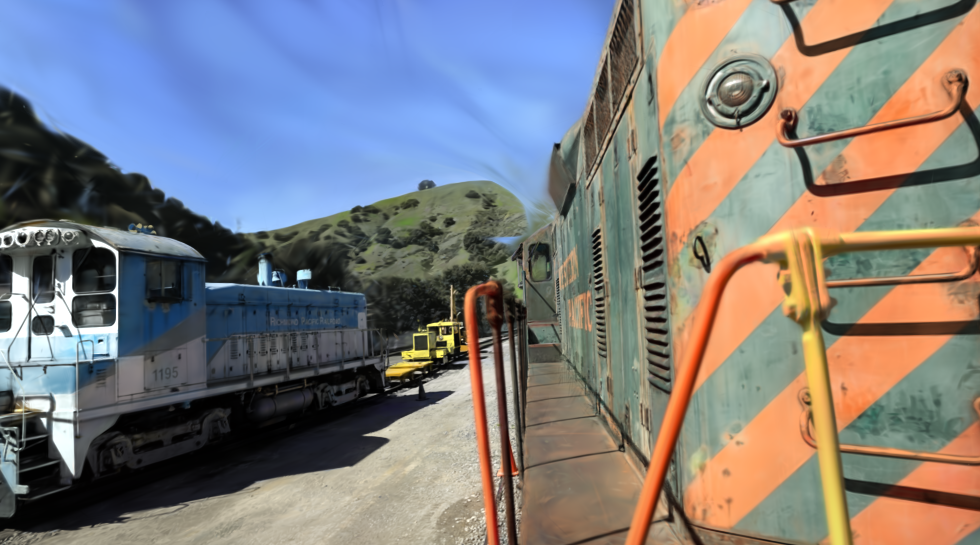} \\

        {\rotatebox{90}{\,\,\, Room}}&
        \includegraphics[width=0.98\linewidth]{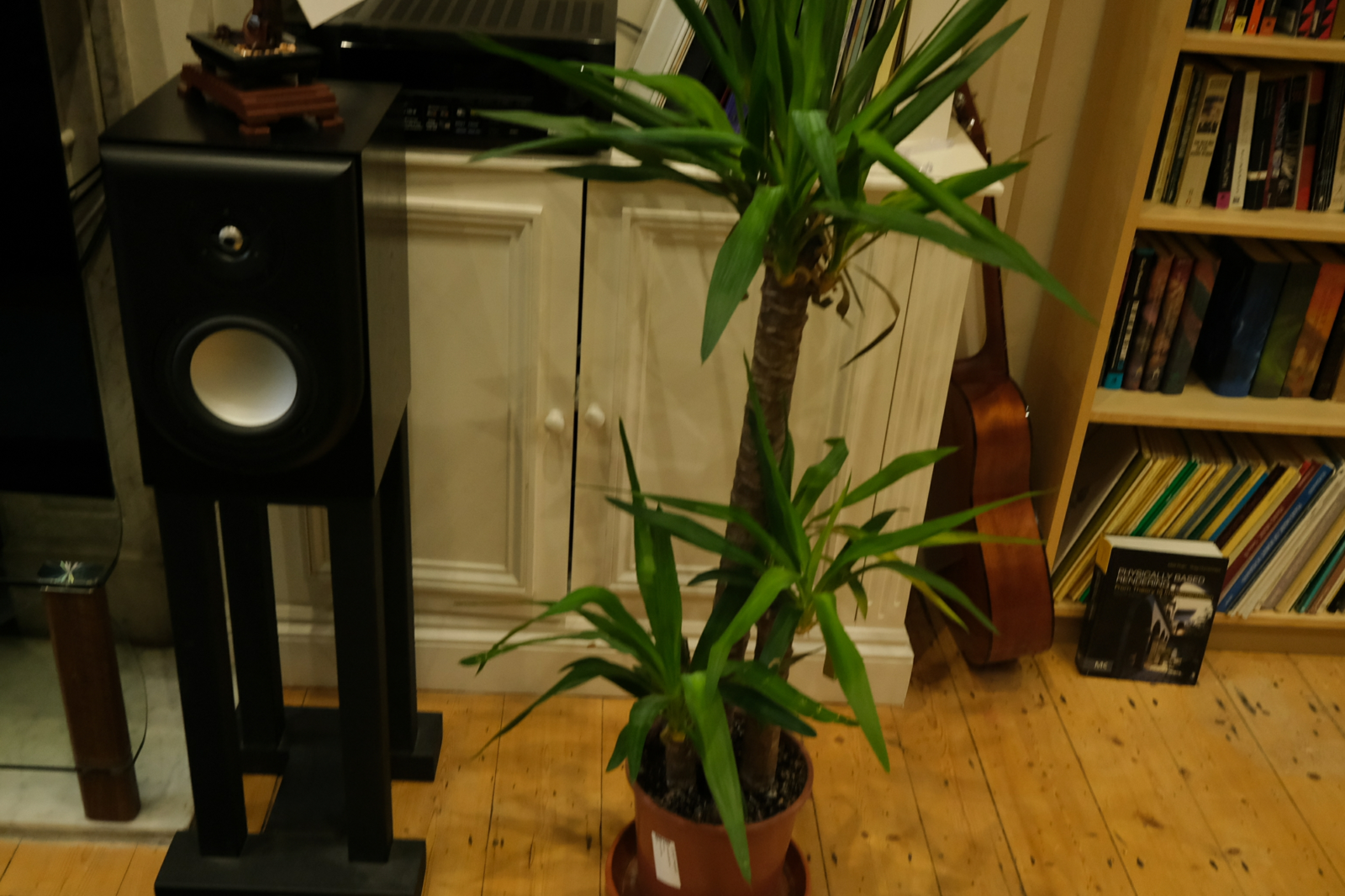} &
        \includegraphics[width=0.98\linewidth]{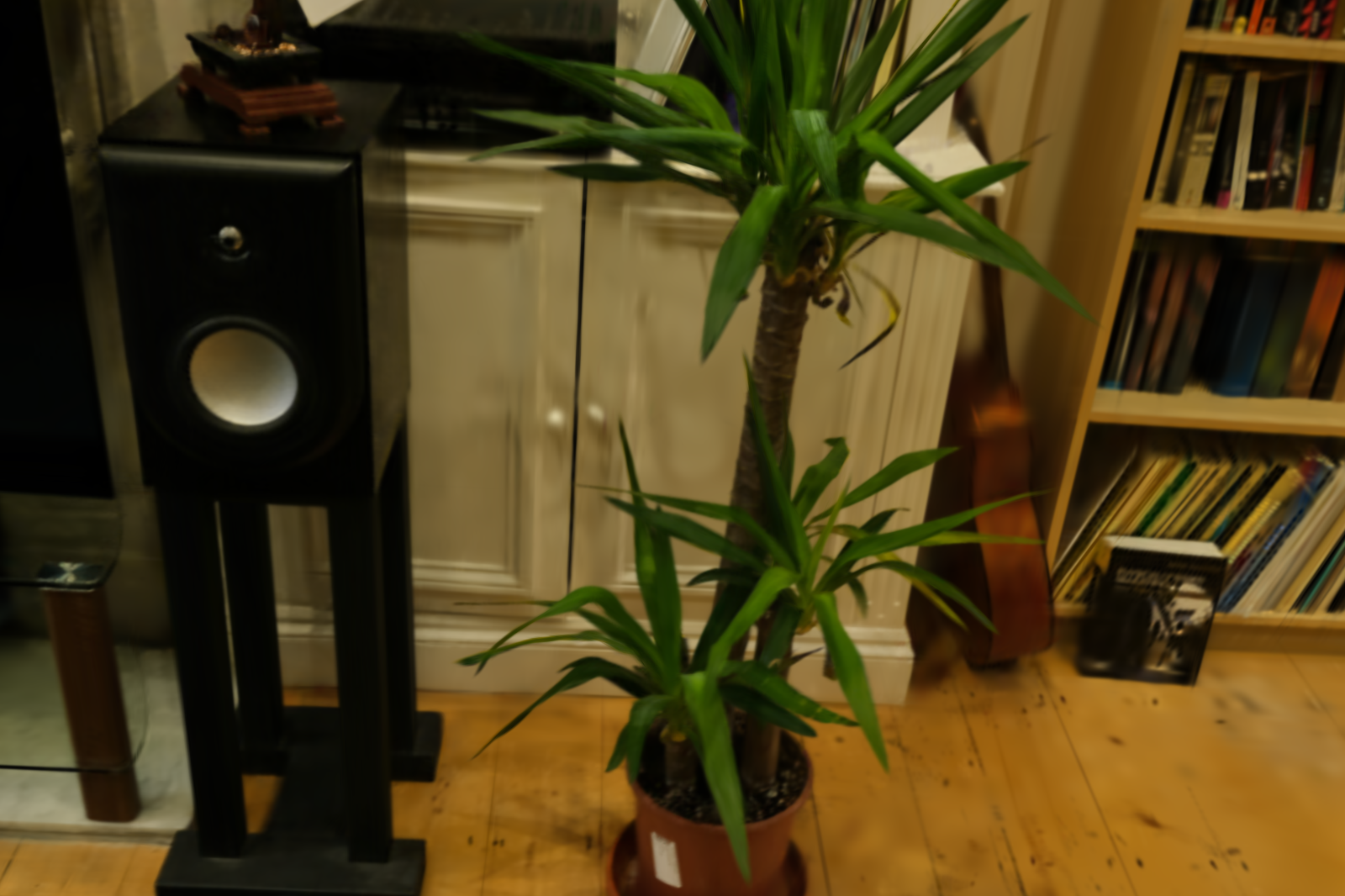} \\
       
    \end{tabularx}
    \makeatletter\def\@captype{figure}\makeatother
    \caption{Qualitative results of \X compared to ground truth reconstruction}
    \vspace{-10pt}
\end{figure}

\end{document}